\documentclass[apj]{emulateapj}

\shorttitle{Torus geometry difference between polarized and non-polarized BLR AGN}
\shortauthors{Ichikawa et al.}

\usepackage{amsmath}
\usepackage{color}
\usepackage{url}
\usepackage{ulem}
\usepackage{multirow}
\usepackage{longtable}

\usepackage{graphicx}
\usepackage{epstopdf}

\def\B{\textsc{BayesClumpy}}
\def\R{\textsc{RedCan}}

\begin{document}


\title{On the difference of torus geometry between hidden and non-hidden broad line active galactic nuclei}


\author{Kohei~Ichikawa\altaffilmark{1,14}, 
Christopher~Packham\altaffilmark{2},
Cristina~Ramos~Almeida\altaffilmark{3,4,15},
Andr\'{e}s~Asensio~Ramos\altaffilmark{3,4},
Almudena~Alonso-Herrero\altaffilmark{5,16},
Omaira~Gonz\'{a}lez-Mart\'{i}n\altaffilmark{3,4},
Enrique~Lopez-Rodriguez\altaffilmark{2},
Yoshihiro~Ueda\altaffilmark{1},
Tanio~D\'{i}az-Santos\altaffilmark{6,7},
Moshe~Elitzur\altaffilmark{8},
Sebastian~F.~H\"{o}nig\altaffilmark{9,15}, 
Masatoshi~Imanishi\altaffilmark{10}, 
Nancy~A.~Levenson\altaffilmark{11},
Rachel~E.~Mason\altaffilmark{12},
Eric~S.~Perlman\altaffilmark{13},
and
Crystal~D.~Alsip\altaffilmark{2}
}
\affil{\altaffilmark{1}
Department of Astronomy, Kyoto University, Kitashirakawa-Oiwake-cho, Sakyo-ku, Kyoto 606-8502, Japan}
\affil{\altaffilmark{2}
Department of Physics and Astronomy, University of Texas at San Antonio, One UTSA Circle, San Antonio, TX 78249, USA}
\affil{\altaffilmark{3}
Instituto de Astrof\'{i}sica de Canarias, C/V\'{i}a L\'{a}ctea, s/n, E-38205 La Laguna, Tenerife, Spain}
\affil{\altaffilmark{4}
Departamento de Astrof\'{i}sica, Universidad de La Laguna, E-38205 La Laguna, Tenerife, Spain}
\affil{\altaffilmark{5}
Instituto de F\'{i}sica de Cantabria, CSIC-Universidad de Cantabria, 39005 Santander, Spain}
\affil{\altaffilmark{6}
Spitzer Science Center, California Institute of Technology, MS 220-6, Pasadena, CA 91125, USA}
\affil{\altaffilmark{7}
Nucleo de Astronomia de la Facultad de Ingenieria, Universidad Diego Portales, Av. Ejercito Libertador 441, Santiago, Chile}
\affil{\altaffilmark{8}
Department of Physics and Astronomy, University of Kentucky, Lexington, KY 40506-0055, USA}
\affil{\altaffilmark{9}
School of Physics and Astronomy, University of Southampton, Southampton SO17 1BJ, UK}
\affil{\altaffilmark{10}
Subaru Telescope, 650 North A'ohoku Place, Hilo, HI 96720, USA}
\affil{\altaffilmark{11}
Gemini Observatory, Southern Operations Center, c/o AURA, Casilla 603, La Serena, Chile}
\affil{\altaffilmark{12}
Gemini Observatory, Northern Operations Center, 670 N. A'ohoku Place, Hilo, HI 96720, USA}
\affil{\altaffilmark{13}
Department of Physics and Space Sciences, 150 W. University Blvd., Florida Institute of Technology, Melbourne, FL 32901, USA}

\email{ichikawa@kusastro.kyoto-u.ac.jp}


\altaffiltext{14}{JSPS research fellow (DC1), Japan}
\altaffiltext{15}{Marie Curie fellow}
\altaffiltext{16}{Visiting Professor, Department of Physics and Astronomy, University of Texas at San Antonio}


\begin{abstract}
We present results from the fitting of infrared (IR) spectral energy distributions of 21
 active galactic nuclei (AGN) with clumpy torus models. 
 We compiled high spatial resolution ($\sim 0.3$--$0.7$ arcsec)
  mid-IR $N$-band spectroscopy, $Q$-band imaging and 
  nuclear near- and mid-IR photometry from the literature. 
  Combining these nuclear near- and mid-IR observations, 
  far-IR photometry and clumpy torus models, 
  enables us to put constraints on the torus properties and geometry. 
We divide the sample into three types according to the broad line region (BLR) properties;
type-1s, type-2s with scattered or hidden broad line region (HBLR) previously
 observed, and type-2s without any published HBLR signature (NHBLR).
Comparing the torus model parameters gives us the first quantitative torus geometrical view for each subgroup. 
We find that NHBLR AGN have smaller torus opening angles 
and larger covering factors than those of HBLR AGN. 
This suggests that the chance to observe scattered (polarized) flux from the BLR in NHBLR could be reduced
by the dual effects of 
  (a) less scattering medium due to the reduced scattering volume given the small torus opening angle and 
  (b) the increased torus obscuration between the observer and the scattering region. 
  These effects give a reasonable explanation for the lack of observed HBLR in some type-2 AGN.
\end{abstract}





\section{INTRODUCTION}
While active galactic nuclei (AGN) present a variety of observational characteristics,
 the unified model for AGN proposes the ubiquitous presence of an obscuring torus
  around their central engines, and that all AGN are fundamentally the same \citep{ant93}.
This optically and geometrically thick torus produces the effect of a line of sight viewing
 angle dependency.
Type-1 AGN are observed with the direct view of fast moving material close to the 
supermassive black hole (SMBH), resulting in broad emission lines in their spectra, 
while type-2 AGN are observed from an edge-on view and the torus blocks the broad
 emission line region (BLR) component from our line of sight.
%
%
The most compelling evidence of the unified model was the detection of polarized
 broad emission lines (PBLs) in type-2 AGN \citep[e.g.,][]{ant85}. 
Further evidence supporting the unified model comes from infrared (IR) observations
of several type-2 AGN showing the existence of obscured/hidden broad line
 regions (HBLRs) detectable only with dust penetrating infrared observations
  \citep[e.g.,][]{bla90,nag02,reu03,ram08}.

Against the fact that the observations generally support the unified model,
there is the question of why some, but not all, type-2 AGN do not show any
 observational signs of PBLs.
\cite{tra01,tra03} and \cite{mor01} found that only 30--50\% of type-2 AGN show PBLs.
Some studies have advocated that the non-detection of a PBL is due to genuine
 lack of a BLR \citep[e.g.,][]{tra11}.
Others have suggested that the non-detection is due to obscuration effects,
 rendering the detection of PBLs as difficult or impossible, even with deep
  near-IR (NIR) spectro-polarimetric observations \citep{ale01}.
 Using a statistically complete \textit{IRAS} 60~$\mu$m selected type-2 AGN catalog,
\cite{hei97} investigated the relationship between the detectability of PBLs and
 IR color as an indicator of the torus inclination angle. They showed that only AGN with a low torus inclination angle
  have high detection rate of PBLs compared to those with high inclinations.
This result strongly suggests that PBLs could be obscured
when there is an edge-on view through the torus and/or nuclear obscuration in the host galaxies.
In addition to the optical spectro-polarimetry, X-ray observations suggest that
 there is a weak evidence showing different absorption in two types of type-2 AGN. 
 \cite{gu01} found that the AGN with PBL have slightly lower column density ($N_{\rm H}$) 
 than those without PBL.
Similarly, \cite{lum04} showed that the detection rate of PBL decreases
   as a function of $N_{\rm H}$, suggesting that the absorption effect by dusty torus could play
    a role on the detectability of PBL in AGN.

To understand the role of the obscuration by the torus in type-2 AGN 
and the detectability of PBLs, knowing the torus geometry
 and properties is crucial. 
In recent years much progress has been made toward understanding
 the torus geometrical structure.
Thanks to the improvement of computing power, more physically realistic
 torus models assuming ``clumpy'' distributions (called clumpy torus models) have been coded by several authors
  \citep{nen02,nen08a,nen08b,hon06,sch08,hon10,sta12}. 
 These models readily reproduce high spatial resolution nuclear NIR to mid-IR (MIR) 
 spectral energy distributions (SEDs) and spectra of AGN with a compact torus 
 of $<10$~pc radius \citep[e.g.,][]{ram09,ram11,nik09,alo11,hon11,lir13}.
On the other hand, traditional smooth torus models
  \citep{pie92,pie93,efs95} had difficulties to describe the variety of nuclear
   SEDs of nearby AGN \citep[e.g.,][]{alo03,gan09, asm11, ich12}.
Still, the true torus morphology remains far from being conclusively
 determined until future observations resolve the torus. 
 Both smooth and clumpy torus models have degeneracies, 
 and from SED-fitting alone is not possible to choose one or the other
 \citep[see][ for a comparison between smooth and clumpy torus models]{fel12}.

In this paper, under the assumption that the torus follows
 a clumpy distribution of dust, which we consider more realistic in principle
 , we discuss how the precise modeled torus morphology plays 
 a key role in the probability of the detection of PBL by fitting clumpy
  torus models to our series of IR SEDs.
  We use 21 high spatial resolution MIR spectra in combination 
  with NIR to far-IR (FIR) photometry, constituting one of the largest
   compilations of nuclear IR SEDs of AGN in the local universe. 
These SEDs afforded by 8~m class telescopes
 minimize contamination of  the MIR torus spectra from surrounded diffuse
  MIR emission from warm dust and/or starformation region emission, and hence we are
   able to construct the highest fidelity torus SED.
Therefore we can probe the detectability of the PBL with the least amount
 of host galaxy contaminations than ever before.
 The luminosity dependence of the torus morphology will be discussed
  in a forthcoming paper (Ichikawa~et~al.~in~prep.).

\setlength{\tabcolsep}{0.05in} 
\begin{deluxetable*}{cc*{15}c}
\tablecaption{Properties of the Sample
\label{tbl-1}}
\tablewidth{0pt}
\tablehead{
\colhead{Name} & \colhead{$z$} & \colhead{$d$} & \colhead{Slit/Size} &
 \colhead{Type} & \colhead{Group} & \colhead{$N_{\rm H}$} & \colhead{$\log L_{\rm bol}^{\rm (lit)}$} &
 \colhead{$b/a$} &   \colhead{$A_{\rm V}$} & \colhead{$i$} & \colhead{Ref}
 \\
\colhead{(1)} & \colhead{(2)} & \colhead{(3)} & \colhead{(4)} &
\colhead{(5)} & \colhead{(6)} & \colhead{(7)} & \colhead{(8)} &
\colhead{(9)} & \colhead{(10)} & \colhead{(11)}  & \colhead{(12)}
}
\startdata
NGC~1365      & 0.0055 &   18 &    0.35/31 &    Sy1.8 &     Type-1 &     23.6 & 42.9 &   0.5 &       $<$5 &  $\cdots$  & (A1,B1,B1,$\cdots$)\\
NGC~4151      & 0.0033 &   13 &    0.36/23 &    Sy1.5 &  Type-1 &     22.8 & 43.7 &  0.71 &  $\cdots$  &  $\cdots$  & (A9,A9,$\cdots$,$\cdots$)\\
IC~4329A      &  0.016 &   65 &   0.75/240 &    Sy1.2 &  Type-1 &     21.8 & 43.6 &  0.28 &  $\cdots$  &  $\cdots$  & (A10,A9,$\cdots$,$\cdots$)\\
NGC~7469      &  0.016 &   66 &   0.75/240 &      Sy1 &  Type-1 &     20.7 & 45.1 &  0.72 &  $\cdots$  &  $\cdots$  & (A9,A9,$\cdots$,$\cdots$)\\
\hline
NGC~1068      & 0.0038 &   15 &    0.36/26 &      Sy2 &     HBLR &    $>25$ & 45.0 &  0.85 &  $\cdots$  &     60--90 & (A2,A9,$\cdots$,A9)\\
NGC~2110      & 0.0078 &   31 &    0.36/54 &      Sy2 &     HBLR &     22.5 & 43.9 &  0.74 &          5 &         40 & (A9,A9,A9,A9)\\
MCG~-5-23-16  & 0.0085 &   34 &   0.75/120 &      Sy2 &     HBLR &     22.2 & 44.4 &  0.46 &       $>$6 &         53 & (A9,A9,A9,A9)\\
NGC~3081      &  0.008 &   32 &   0.65/100 &      Sy2 &     HBLR &     23.9 & 43.8 &   0.8 &  $\cdots$  &  $\cdots$  & (A3,B2,$\cdots$,$\cdots$)\\
NGC~3227      & 0.0039 &   17 &    0.75/62 &      Sy2 &     HBLR &     22.2 & 43.4 &  0.68 &  $\cdots$  &  $\cdots$  & (A11,A9,$\cdots$,$\cdots$)\\
Circinus     & 0.0014 &    4 &    0.60/12 &      Sy2 &     HBLR &     24.6 & 43.6 &  0.44 &          9 &     60--90 & (A8,A9,A9,A9)\\
NGC~5506      & 0.0062 &   25 &    0.36/44 &      Sy2 &     HBLR &     22.4 & 44.2 &  0.30 &     $\ge$11 &         40 & (A9,A9,A9,A9)\\
IC~5063       &  0.011 &   46 &   0.67/150 &      Sy2 &     HBLR &     23.3 & 44.5 &  0.68 &          7 &  $\cdots$  & (A2,A9,A9,$\cdots$)\\
NGC~7582      & 0.0053 &   21 &    0.75/76 &      Sy2 &     HBLR &     22.7 & 43.3 &  0.42 &       8,13 &  $\cdots$  & (A9,A9,A9,$\cdots$)\\
NGC~7674      &  0.029 &  118 &   0.75/430 &      Sy2 &     HBLR &    $>$25 &   45.0 &  0.91 &  $\sim$3--5 &  $\cdots$  & (A9,A9,A9,$\cdots$)\\
\hline
NGC~1386      & 0.0029 &   11 &    0.31/17 &      Sy2 &    NHBLR &  $>$25.0 & 42.9 &   0.4 &  $\cdots$  &      65,85 & (A2,B2,$\cdots$,C1)\\
NGC~3281      &  0.011 &   43 &    0.35/73 &      Sy2 &    NHBLR &     24.3 & 44.6 &   0.4 &  $\cdots$  &  $\cdots$  & (A4,B1,$\cdots$,$\cdots$)\\
Cen~A         & 0.0018 &    3 &    0.65/11 &      Sy2 &    NHBLR &     23.7 & 44.0 &   0.4 &       14.0 &  $\cdots$  & (A5,B2,A9,$\cdots$)\\
NGC~5135      &  0.014 &   59 &   0.70/200 &      Sy2 &    NHBLR &  $>$25.0 & 44.4 &   0.7 &  $\cdots$  &  $\cdots$  & (A2,B2,$\cdots$,$\cdots$)\\
NGC~5643      &  0.004 &   16 &    0.35/29 &      Sy2 &    NHBLR &     23.8 & 42.7 &   0.9 &  $\cdots$  &  $\cdots$  & (A6,B5,$\cdots$,$\cdots$)\\
NGC~5728      & 0.0094 &   40 &    0.35/69 &      Sy2 &    NHBLR &     23.6 & 44.5 &   0.6 &  $\cdots$  &  $\cdots$  & (A7,B6,$\cdots$,$\cdots$)\\
NGC~7172      & 0.0087 &   35 &    0.36/61 &      Sy2 &    NHBLR &     22.9 & 43.8 &  0.46 &  $\cdots$  &  $\cdots$  & (A2,A9,$\cdots$,$\cdots$)
\enddata
\tablecomments{Sample properties. The sample is divided into three subgroups with type-1/HBLR/NHBLR respectively from top to bottom.
(1) object name; 
(2) redshift; 
(3) luminosity distance (Mpc) gathered from literature for the case of nearby sources. 
Within the sample of \cite{gon13}, for NGC~1365, NGC~1386, NGC~1808, NGC3081,
 NGC~3281, and Cen~A, the values of distance to the galaxies have  been taken from
  \cite{ram09}. For NGC~5643, the distance has been taken from \cite{gua04}. For the sample
  of \cite{alo11}, we gathered them from \cite{alo11}. 
  For the other sources, we calculated the distances by using cosmological parameter $H_{0} = 75$km s$^{-1}$Mpc$^{-1}$;
(4) slit width (arcsec) / physical size (pc); 
(5) Seyfert class of AGN.
(6) Sub group of AGN. Type-1 represents type-1 AGN (Sy 1 to Sy 1.9) based on optical spectroscopy. 
HBLR represents type-2 (Sy2) AGN with hidden broad line region signs, 
and NHBLR represents type-2 AGN without any published hidden broad line regions signs.
(7) hydrogen column density; 
(8) logarithm of bolometric luminosity (erg/s) which is taken from \cite{gon13, alo11}. We use a typical bolometric correction of 20 \citep{elv94}. 
(9) The axial ratio; the ratio of the minor to major axis of the host galaxies. All information is taken from \cite{gon13};
(10) Foreground extinction in the unit of mag; 
(11) inclination angle of the torus. \cite{lev06} derives the viewing angle of accretion disk of NGC~1386 and we here assume that the accretion disk and the torus are located in the same plane;
(12) References of column (6), (7), (10), and (11). ``$\cdots$'' denotes no reference.
\\
\textbf{References.} 
(A1) \cite{alo12}; (A2) \cite{tra01}; (A3) \cite{mor00}; (A4) \cite{nic03}; (A5) \cite{ale99}; (A6) \cite{gu01};
 (A7) \cite{tra03}; (A8) \cite{wan07}; (A9) \cite{alo11}; (A10) \cite{ver06}; (A11) \cite{ima02};
(B1) \cite{tue08}; (B2) \cite{mar12}; (B3) \cite{bri11}; (B4) \cite{ito08}; (B5) \cite{gua04}; (B6) \cite{gou12};
(C1) \cite{lev06}
}\\
\end{deluxetable*}

 \section{OBSERVATIONS}

\subsection{The Sample}
Our principal motivation in this study is to investigate whether the torus model morphology plays a major role in the chance of PBL detections.
To achieve this goal, we compiled the nearby AGN sources from the MIR samples of \cite{gon13} (21 sources) and \cite{alo11} (13 sources)
as both samples already compiled the currently available data set of ground-based $N$ band spectroscopy.
Of the 34 sources, five are in common in both samples, therefore the total number is 29 sources.
We further set the criteria for survey inclusion that the objects must have at least one high spatial resolution NIR (1--5~$\mu$m)
measurement, as the NIR bands significantly help to constrain the torus parameters \citep{ram14}.
22 out of 29 sources fulfilled this criterion.
We also removed NGC~1808 from this study as controversy remains to whether it hosts
AGN or ultra-luminous X-ray sources in the galactic center due to the low X-ray luminosity $\log L_{\rm 2-10~keV} = 40.4$~erg~s$^{-1}$
\citep{sca93,jim05}.
We summarize the properties of the 21 sources in Table~1.

Our sample spans AGN bolometric luminosities taken from the literature ($L_{\rm bol}^{(\rm lit)}$) in the range 
$\log L_{\rm bol}^{(\rm lit)} =$42.7--45.1~erg~s$^{-1}$ (see Table~1), with a mean value of 44.0~erg~s$^{-1}$. 
This value is fairly consistent with that of magnitude-limited Seyfert catalogs
 \citep{mai95,ho97}. This suggests that our sample could be
   representative of AGN and their tori in the local universe, although the sample is not complete.

\subsection{New Observations}
 We obtained $N$ (Si2 filter; the central wavelength with $\lambda_{\rm c}=8.73$~$\mu$m
  and 50\% cutoff range of $\Delta\lambda=0.39$~$\mu$m) 
and $Q$ (Qa filter; $\lambda_{\rm c}=18.06$~$\mu$m and $\Delta\lambda=0.76$~$\mu$m)
 band imaging data of NGC~5135 and
 NGC~5643,  observed by T-ReCS (Program ID GS-2012A-Q-43, PI: Nancy Levenson).
The standard MIR chop-nod technique was performed for the observations. 
The data reduction was made by using \R\ \citep{gon13}. 

\subsection{Published Data from the Literature}
 We collected the estimated values of the nuclear NIR to MIR emission when available.
  We compiled nuclear NIR data from both ground- and space-based telescopes 
  such as VLT/NACO and \textit{HST}/NICMOS. 
  The only exception is the galaxy NGC 5728, whose only NIR flux is from 
  2MASS \citep{pen06}, which we use as an upper limit. 
  Information about the NIR fluxes used here is compiled in Table~2 (columns 2-6). 

In order to further reduce the parameters space at longer wavelengths, 
we used  the \textit{Spitzer}/IRS 30~$\mu$m continuum fluxes reported
 by \cite{deo09} as upper limits in our fits. 
 We do that because the star formation component can be important at 
 20--30~$\mu$m, and beyond 30~$\mu$m completely overwhelms the
  AGN torus emission in most cases \citep{net07} at the spatial resolutions
   afforded by \textit{Spitzer}. 
   We only consider the \textit{Spitzer} 30~$\mu$m photometry as data points,
    and also included the \textit{Spitzer}/IRS spectroscopy when the AGN spectral
     turnover at 20--30~$\mu$m is clearly seen in the IRS spectra. 
     This feature suggests that the torus emission is dominant even in the large
      aperture data from \textit{Spitzer} \citep{alo12b,hon14}.
       Only two sources fulfilled the criterion (IC~4329A and MCG~-5-23-16).
For those galaxies, we also collected available \textit{Herschel}/PACS data.
All the FIR flux information is tabulated in column 9 to 11 in Table~2.

The errors were estimated using the prescription given by \cite{alo12}.
For the NACO AO observations, 
we used 20\% in $J$ band and 15\% in the $HKLM$ band. 
For the other ground-based observation data, we applied 30\% for $J$ band, 25\% for $H$
 and $K$ band, and 20\% for the $L$ band. 
 These errors include the photometric error, the background subtraction uncertainty, 
 and the uncertainty from estimating the unresolved flux. $M$ band fluxes were always
  used as upper limits due to the difficulties of estimating the unresolved component. 
  For the NICMOS observations, we used 20\% for the $J$ band, 20\% for the $H$ and $K$ band.
For the $N$ and $Q$ band,we use 15\% and 25\% errors, respectively.

\setlength{\tabcolsep}{0.022in} 
\begin{deluxetable*}{cc*{11}c}
\tablecaption{List of Photometry
\label{tbl-1}}
\tablewidth{0pt}
\tablehead{
\colhead{Name} & \colhead{$J$} & \colhead{$H$} & \colhead{$K$} & 
\colhead{$L$} & \colhead{$M$} & \colhead{$N$} & \colhead{$Q$}& 
\colhead{$30~\mu$m} & \colhead{$70~\mu$m} & \colhead{$160~\mu$m} &
\colhead{Ref}
 \\
\colhead{(1)} & \colhead{(2)} & \colhead{(3)} & \colhead{(4)} &
\colhead{(5)} & \colhead{(6)} & \colhead{(7)} & \colhead{(8)} &
\colhead{(9)} & \colhead{(10)} & \colhead{(11)} & \colhead{(12)}
}
\startdata
\multicolumn{12}{c}{Type-1}\\
\hline
NGC~1365       &   $\cdots$         & $8.3\pm0.83$    &    $\cdots$    &       $\cdots$    &         $\cdots$    & $203\pm30.0$          & $818\pm204$           & $<12.3$   &    $\cdots$    &       $\cdots$    & (A1,B1)\\ 
NGC~4151        & $69\pm14$     &  $104\pm10.4$  & $178\pm17.8$& $<325$             &          $\cdots$    & $1320\pm198$        & $3200\pm800$         & $<3.64$   & $\cdots$    &   $\cdots$    & (A7,A7) \\
IC~4329A          &      $\cdots$      & $50.0\pm8.0$    & $102\pm10$   & $<210$              &      $\cdots$    & $1014\pm150$         &       $\cdots$    & $1.52\pm0.015$   & $<1.79$ & $<0.97$  & (A7,A7) \\
NGC~7469        & $16\pm3.2$    & $40\pm4.0$        & $68\pm6.8$     & $<84$               &     $\cdots$    & $506\pm76$            &  $1350\pm340$         &  $\cdots$    &   $\cdots$    &   $\cdots$    & (A7,A7)\\ 
\hline
\multicolumn{12}{c}{HBLR}\\
\hline
NGC~1068        & $9.8\pm2.0$  & $98.0\pm15.0$    & $445\pm100$ & $920\pm140$ & $2270\pm340$& $10000\pm1500$ & $21800\pm5400$     &    $\cdots$    &   $\cdots$    &  $\cdots$    &(A7,A7)\\
NGC~2110       &      $\cdots$      &     $\cdots$    &        $\cdots$    & $<33.0$            & $<198$            & $294\pm44$           &  $561\pm140$                       & $<0.8$    &    $\cdots$    &   $\cdots$    & (A7,A7) \\
MCG~-5-23-16 & $1.1\pm0.33$& $3.7\pm0.93$     & $10.7\pm2.7$ & $79.5\pm16.0$& $<139.4$        & $633\pm95$            & $1450\pm360$         &   $\cdots$    & $<1.45$  & $<0.45$    & (A7,A7)\\
NGC~3081       &       $\cdots$     & $0.22\pm 0.04$ & $\cdots$    &      $\cdots$    &          $\cdots$    & $83\pm12.5$            & $231\pm57.8$          & $<1.10$   &      $\cdots$    &           $\cdots$    & (A2,B1) \\
NGC~3227        &       $\cdots$    & $11\pm1.1$       & $23\pm2.3$     & $<47.0$            &     $\cdots$    & $320\pm48$             & $1100\pm275$                    & $<1.76$   &  $\cdots$    &  $\cdots$    & (A7,A7)\\ 
Circinus              & $<1.60$          & $4.77\pm0.72$   & $19\pm2.9$     & $380\pm57$   & $1900\pm285$&$5600\pm840$       &  $12800\pm3200$    &     $\cdots$    &  $\cdots$    &    $\cdots$    & (A7,A7)\\
NGC~5506      &  $13\pm3.0$    & $53\pm8.0$         & $80\pm12$     & $290\pm44$    & $<530$            & $900\pm135$         &  $2200\pm550$         & $<4.05$  &    $\cdots$    &        $\cdots$    & (A7,A7) \\
IC~5063             &       $\cdots$    & $0.3\pm0.1$        & $4.8\pm1.0$    &    $\cdots$    &       $\cdots$    & $925\pm139$         &         $\cdots$    & $<3.89$   &    $\cdots$    &   $\cdots$    &(A7,A7) \\
NGC~7582      &         $\cdots$    & $11.0\pm1.6$     & $18.0\pm2.7$  & $96.0\pm14.4$& $141\pm21$  & $384\pm57$           & $527\pm132$            &    $\cdots$    &   $\cdots$    &   $\cdots$    & (A7,A7)\\
NGC~7674      & $1.25\pm0.25$& $5.0\pm0.5$       & $12.3\pm3.1$  & $53.0\pm11.0$& $<108$           & $518\pm78$           &           $\cdots$    & $<1.83$    &   $\cdots$    &    $\cdots$    & (A7,A7)\\
\hline
\multicolumn{12}{c}{NHBLR}\\
\hline
NGC~1386       &         $\cdots$    & $0.2\pm0.04$    &      $\cdots$    &      $\cdots$    &   $\cdots$    & $147\pm22.1$          & $457\pm114$           & $<1.58$   &  $\cdots$    &   $\cdots$    & (A2,B1)  \\
NGC~3281       &      $\cdots$    & $1.3\pm0.33$    & $7.7\pm1.93$ & $103\pm20.6$  & $<207$           & $355\pm53.3$          & $1110\pm278$        &   $\cdots$    &  $\cdots$    &     $\cdots$    &  (A4,B1) \\
Cen~A               & $1.3\pm0.26$ & $4.5\pm0.68$    & $34\pm5.1$     & $200\pm30$     &    $\cdots$    & $710\pm107$           & $2630\pm658$        &   $\cdots$    &   $\cdots$    &     $\cdots$    & (A5,B1) \\
NGC~5135       & $<0.72$           & $0.66\pm0.07$  &     $\cdots$    &       $\cdots$    &        $\cdots$    &$56.36\pm8.454$     & $218.92\pm54.73$ & $<3.03$   &         $\cdots$    &    $\cdots$    &(A2;A6, B2)   \\
NGC~5643       &     $\cdots$    & $<1.7$                 &      $\cdots$    &     $\cdots$    &           $\cdots$    & $101.31\pm15.17$  & $883.2\pm 220.8$   &    $\cdots$    &   $\cdots$    &  $\cdots$    & (A2,B2) \\
NGC~5728       &     $\cdots$    &       $\cdots$    & $<7.1$             &         $\cdots$    &         $\cdots$    & $25\pm3.75$            & $184\pm46$              &  $\cdots$    &     $\cdots$    &   $\cdots$    & (A6,B1) \\\
NGC~7172        &       $\cdots$    & $<0.4$                   & $3.4\pm0.86$ & $30\pm6$        & $<61.4$            & $165\pm30$           &              $\cdots$    & $<0.98$   &     $\cdots$    &     $\cdots$    & (A7,A7) 
\enddata
\tablecomments{
NIR to FIR fluxes used as inputs for \B. 
Units are in mJy for columns 2 to 8, and in Jy for columns 9 to 11. 
Column 9 corresponds to \textit{Spitzer}/IRS 30~$\mu$m continuum fluxes from \cite{deo09},
 which are used as upper limits in the fits. 
 Columns 10 and 11 list the \textit{Herschel}/PACS photometry from \cite{mel14} and are also used as upper limits.
 ``$\cdots$'' represents no flux information.
The references from NIR to MIR band fluxes are tabulated at column~12.\\
\textbf{References.} (A1) \cite{car02} (A2) \cite{qui01}; (A3) \cite{gal08}; (A4) \cite{sim98}; (A5) \cite{mei07}; (A6) \cite{pen06}; (A7) \cite{alo11};
 (B1) \cite{ram09}; (B2) This work.
}\\
\end{deluxetable*}

\subsection{Subsample}
To examine the torus model properties of different AGN populations, we divide the sample
 into subgroups based on whether or not the source has HBLR signs in previously published observations. 
 We first divide the sample into type-1 and type-2 AGN. 
 Although Seyfert 1.8/1.9 are very ambiguous objects \citep[e.g., see][for the details]{eli14},
 here we define type-1 as AGN which have at least one broad emission lines in their optical spectra.
 Therefore, we consider Seyfert 1 to 1.9 as type-1 AGN and Seyfert 2 as type-2 AGN.
 Next, we divide the type-2 AGN into those with any published polarized BLR detections
  in the optical and/or in the NIR (HBLR) and those without (non-HBLR; hereafter NHBLR).
 We use \cite{mar14}, who compiled almost all the previously published polarization information of nearby AGN.
 These spectro-polarimetric data are taken from several large surveys including the
 infrared-selected sample of \cite{hei97}, the FIR flux limited sample of \cite{lum01},
 the distance-limited sample of \cite{mor00,mor02}, and the heterogeneous optical- and
 MIR selected sample of \cite{tra01,tra03}. Mainly the spectro-polarimetric observations were conducted
 with small or medium size telescopes (up to 4~m-class), while only NGC~3081 has been confirmed to have HBLR features with the Keck 10~m telescope \citep{mor00}.
 Therefore, we should note that some HBLR AGN could contaminate the subgroup of NHBLR in the cases where the BLR
 is below the signal-to-noise afforded by the 4~m class telescope observations (see Ramos~Almeida~et~al.,~in~prep.).
 Some sources have currently no published spectro-polarimetric data, but have clear broad emission lines in NIR wavelengths.
 These sources are MCG-5-23-16, NGC~2110, and NGC~7582 \citep{alo11}. All the references used for dividing the sample into each subgroup
  are indicated in column 12 in Table~1.
  
   Finally, the sources in this study are categorized into three groups (type-1, HBLR, and NHBLR; see column~6 in Table~1). 
 The sample contains 4 type-1, 10 HBLR, and 7 NHBLR AGN.

\section{APPLICATION OF TORUS MODEL}
\subsection{Clumpy Torus Model}

We fit the clumpy torus models of \cite{nen08a}, known as \textit{CLUMPY}, to the data using
a Bayesian approach (\B; \citealt{ase09}).
Here we describe the six free \textit{CLUMPY} model parameters used for the SED fitting and the model set-up,
which are listed in Table~3.
The torus clumps are distributed in a smooth, rather than sharp, toroidal-shaped boundary of angular width $\sigma$.
The inner radius ($r_{\rm in}$) of the torus is set by the location of the dust at the sublimation temperature 
($T_{\rm sub} \sim 1500$ K). This is computed using the AGN bolometric luminosity $L_{\rm bol}({\rm AGN})$
\begin{align}
r_{\rm in} = 0.4  \left( \frac{L_{\rm bol}({\rm AGN})}{10^{45}~{\rm erg~s}^{-1}}\right)^{0.5}~{\rm pc}.
\end{align}
The torus has a radial extent ($Y$) defined by $Y=r_{\rm out} / r_{\rm in}$, where $r_{\rm out}$ is the outer radius of the torus. 
%
The average number of clouds along the line of sight ($N_{\rm LOS}$) at a viewing angle $i$ is set as
\begin{align}
N_{\rm LOS} = N_{0} \exp\left[ -\frac{(90-i)^2}{\sigma^2} \right],
\end{align}
where $N_0$ is the average number of clouds along the radial equatorial ray. 
$N_{\rm LOS}$ allows us to derive the escape probability of photons from the AGN central engines ($P_{\rm esc}$).
In the \textit{CLUMPY} dust distribution, the classification of type-1 or type-2 AGN depends on whether or not there
is a clump along the line of sight, which is a function of the viewing angle of the torus, the number of clumps and 
the torus width. This is different from smooth
torus models, for which the classification of an AGN as type-1 or type-2 is solely determined by the viewing angle.
The escape probability of photons passing through the torus at a given
viewing angle ($i$) can be calculated as
\begin{align}
P_{\rm esc} \sim e^{-N_{\rm LOS}}.
\end{align}
In the \textit{CLUMPY} model, the radiative transfer equations are solved
for each clump and thus the calculations depend on the clump distribution
within the torus, the optical depth of each clump, and also its dust composition.
Here we assume each clump has the same optical depth ($\tau_{V}$), 
which is defined at the optical $V$ band. 
The \textit{CLUMPY} model applies a standard cold oxygen-rich interstellar medium dust, 
which is called OHMc dust \citep{oss92}.
The torus clumps are distributed as a power law with index $q$ as a function of radius, $N(r) \propto r^{-q}$.

In addition to these six physical parameters, we add two additional parameters to be fitted or fixed. 
The first parameter is the foreground extinction ($A_V$), unrelated to the torus.
Some authors demonstrated that some AGN have an extremely deep 
 9.7~$\mu$m silicate absorption feature which cannot be reproduced solely by the torus obscuration
 \citep{alo03, pol08,alo11,gou12,gon13}. 
 They suggested that dust in inclined host galaxies can contribute significantly
  to the observed SED and silicate feature absorption. 
 10 out of 21 sources are inclined galaxies with low minor-to-major axis ratios ($b/a \le 0.5$; see Table~1).
 Therefore, some portion of the observed SED is accounted from by cool foreground dust extinction. 
 \cite{alo11} discussed this issue and concluded that for $A_V \ge 5$, the effects of foreground extinction
  cannot be ignored for reproducing the silicate 9.7~$\mu$m feature. 
  We gathered available values of foreground $A_V$ from the literature and compiled them in column~10 in Table~1. 
The other additional parameter accounts for the multiplicative factor that has to be applied 
to match the fluxes of a given model to an observed SED.
Deriving this factor enables us to calculate the model AGN bolometric luminosity
 $L_{\rm bol}^{(\rm mod)}$ \citep{nen08b}. As shown by \cite{alo11}, $L_{\rm bol}^{(\rm mod)}$ 
 reproduces well the values of $L_{\rm bol}^{(\rm lit)}$ for Seyfert galaxies, and therefore, in the 
 following we will refer to $L_{\rm bol}^{(\rm mod)}$ as the bolometric luminosities of the sample studied here.

\begin{deluxetable}{lc*{6}c}
\tabletypesize{\scriptsize}
\tablecaption{Free Parameters of the \B
\label{tbl-4}}
\tablewidth{0pt}
\tablehead{
\colhead{Parameters}
& \colhead{Parameter range} 
}
\startdata
Torus radial thickness ($Y$) & [$5, 30$]\\
Torus angular width ($\sigma$) & [15$^{\circ}, 70^{\circ}$]\\
Number of clouds along an equatorial ray ($N_{0}$) & [$1, 15$]\\
Index of the radial density profile ($q$) & [$0, 3$]\\
Viewing angle ($i$) & [$0^{\circ}, 90^{\circ}$]\\
Optical depth of each cloud ($\tau_{V}$) & [$5, 150$]
\enddata
\tablecomments{Torus radial thickness $Y$ is defined as $Y = r_{\rm out} / r_{\rm in}$, 
where $r_{\rm out}$ is the outer radius and $r_{\rm in}$ is the inner radius. 
The cloud distribution between $r_{\rm out}$ and $r_{\rm in}$ is parameterized as $r^{-q}$.
}\\
\end{deluxetable}


By combining the derived output parameters of the \textit{CLUMPY} model, 
we can derive other important torus morphological parameters as 
the torus outer radius $r_{\rm out}$, defined as:
\begin{align}
r_{\rm out} = r_{\rm in} Y~~{\rm pc}.
\end{align}

We can also calculate the torus scale height $H$ as:
\begin{equation}
H = r_{\rm out} \sin \sigma~~{\rm pc}.
\end{equation}
Finally, we define the ``geometrical'' torus covering factor, 
which is unaffected by the viewing angle, and it is defined by integrating the 
AGN escape probability over all angles \citep{nen08a}. 
 This can be written as
\begin{align}
C_{\rm T} = 1 - \int_{0}^{\pi/2} P_{\rm esc} (\beta) \cos(\beta) d\beta ,
\end{align}
 where $\beta = \pi/2 - i$.
  Considering that our motivation is to characterize the intrinsic torus morphology,  
 the ``geometrical'' torus covering factor is more relevant here than the 
 apparent covering factor.

\begin{figure*}[htbp]
\begin{center}
{\bf \hrulefill}
Type-1
{\bf \hrulefill}
\includegraphics[width=4.6cm]{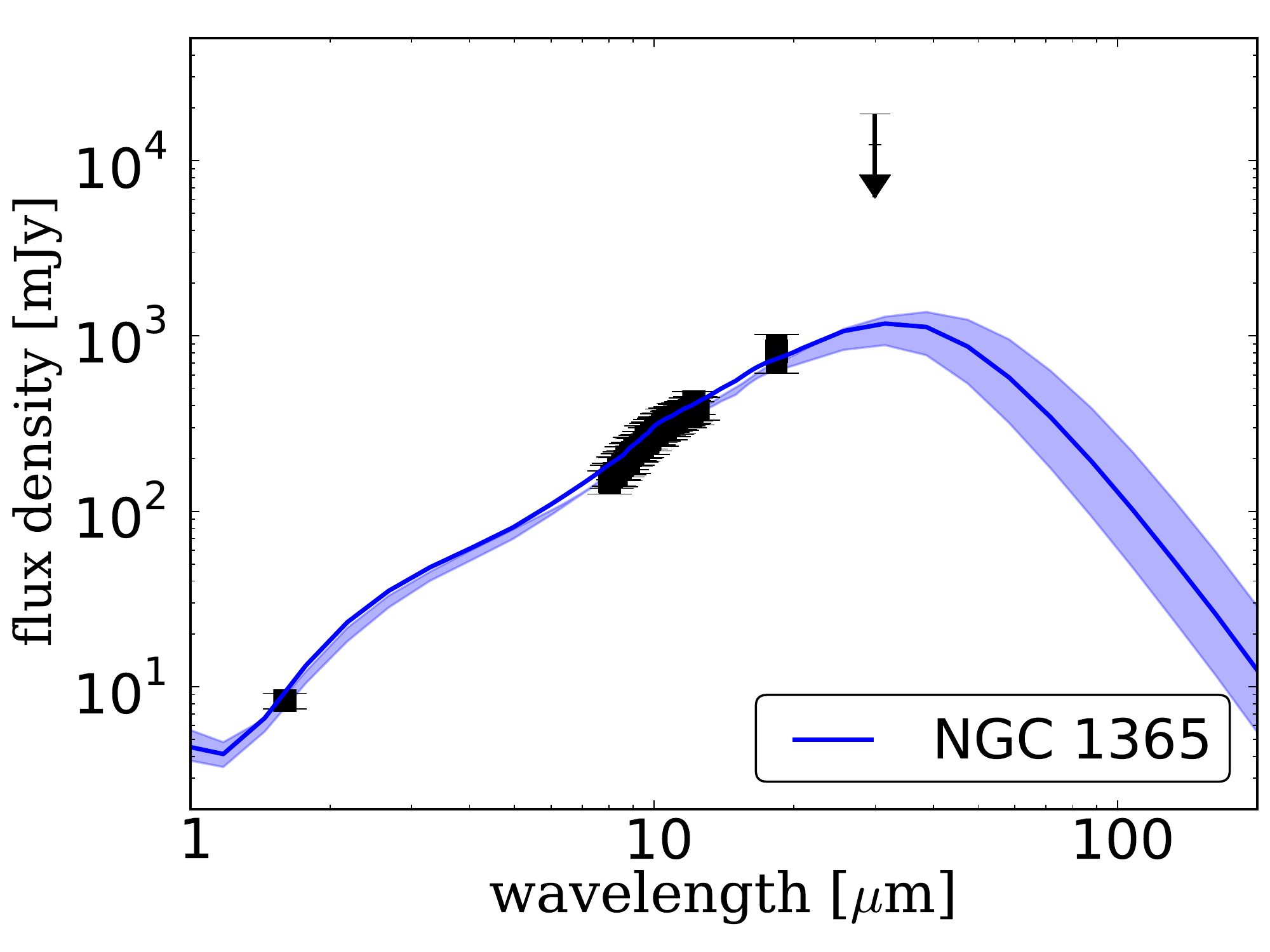}~
\includegraphics[width=4.6cm]{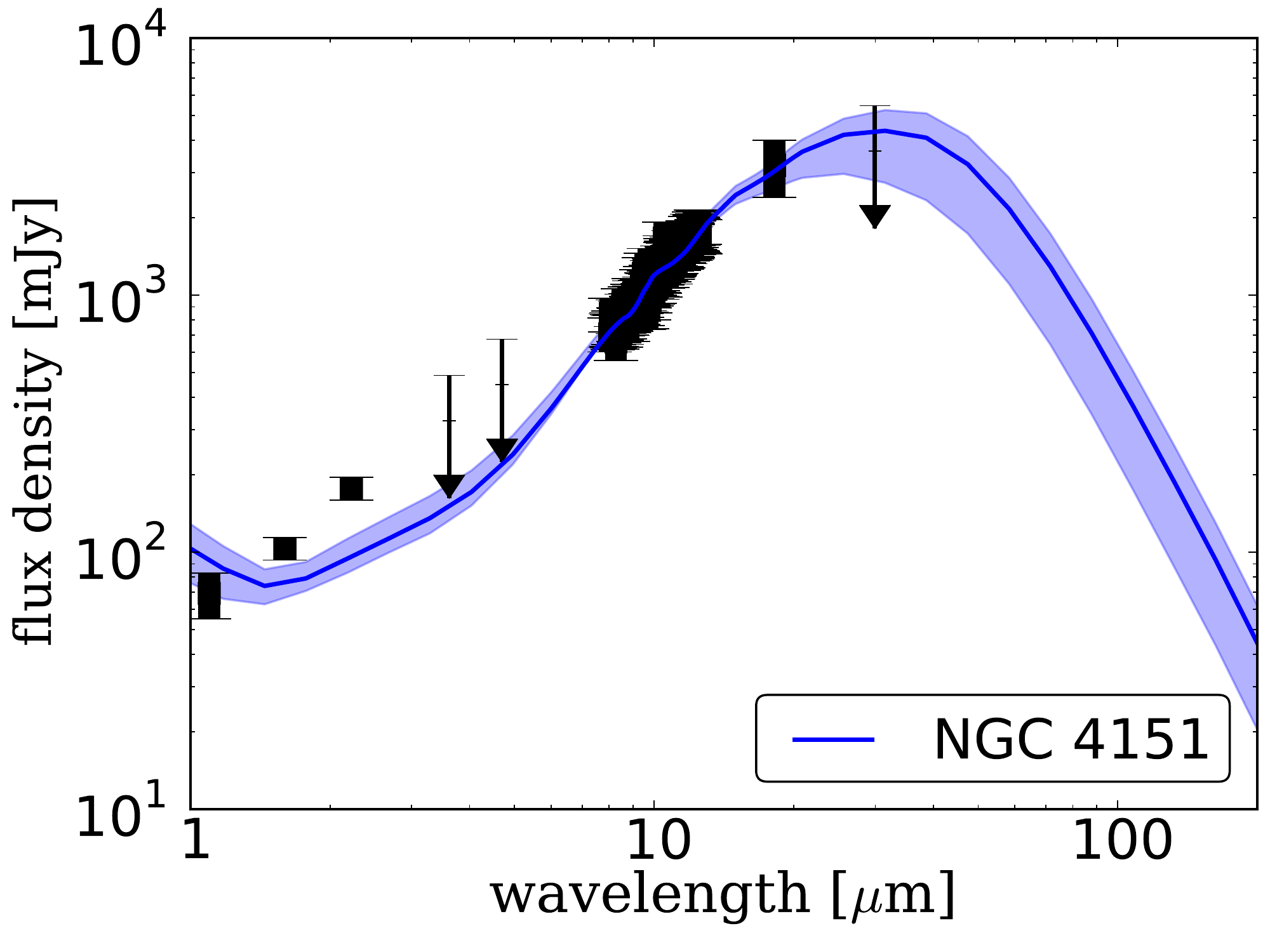}~
\includegraphics[width=4.6cm]{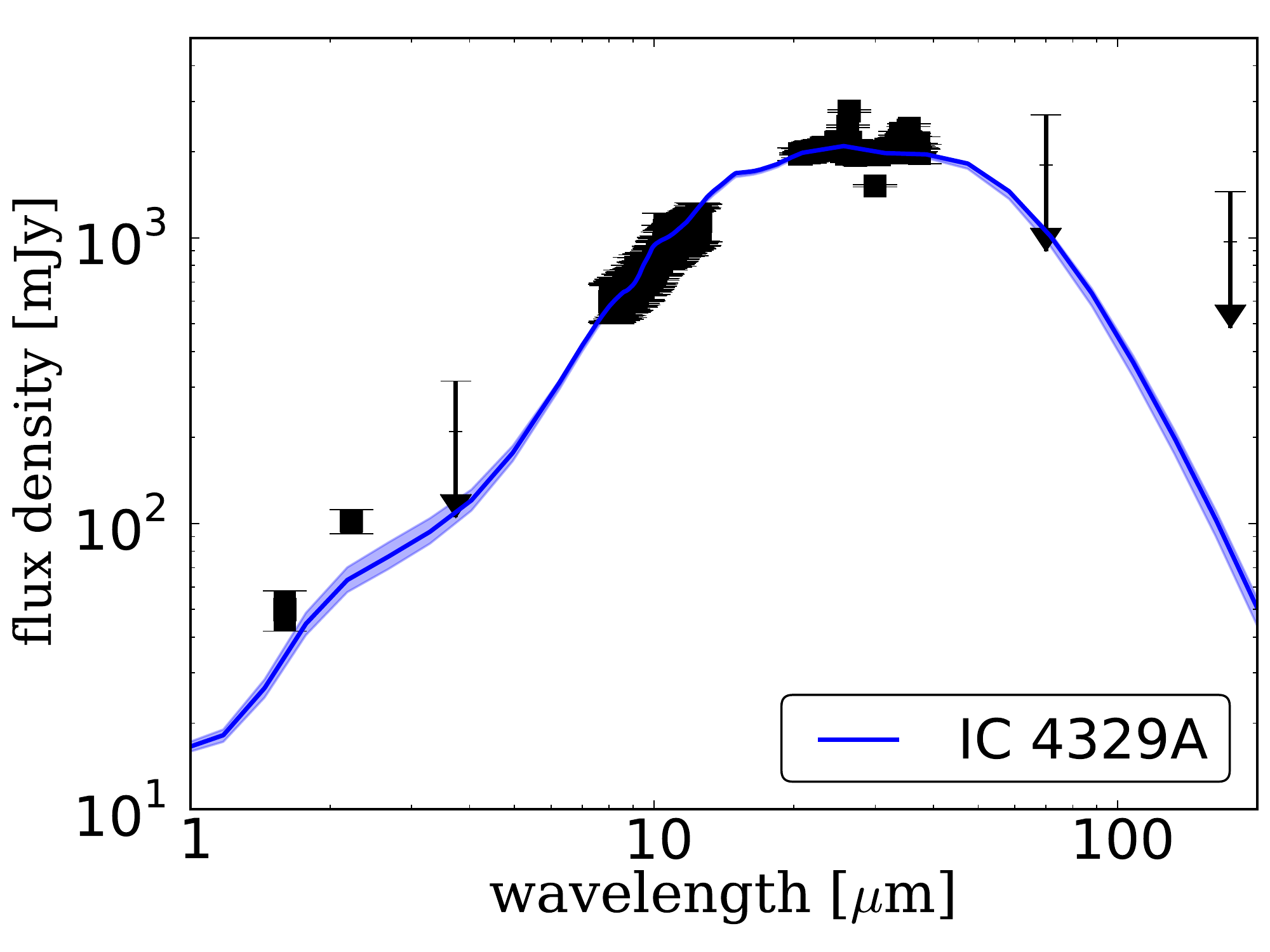}~
\includegraphics[width=4.6cm]{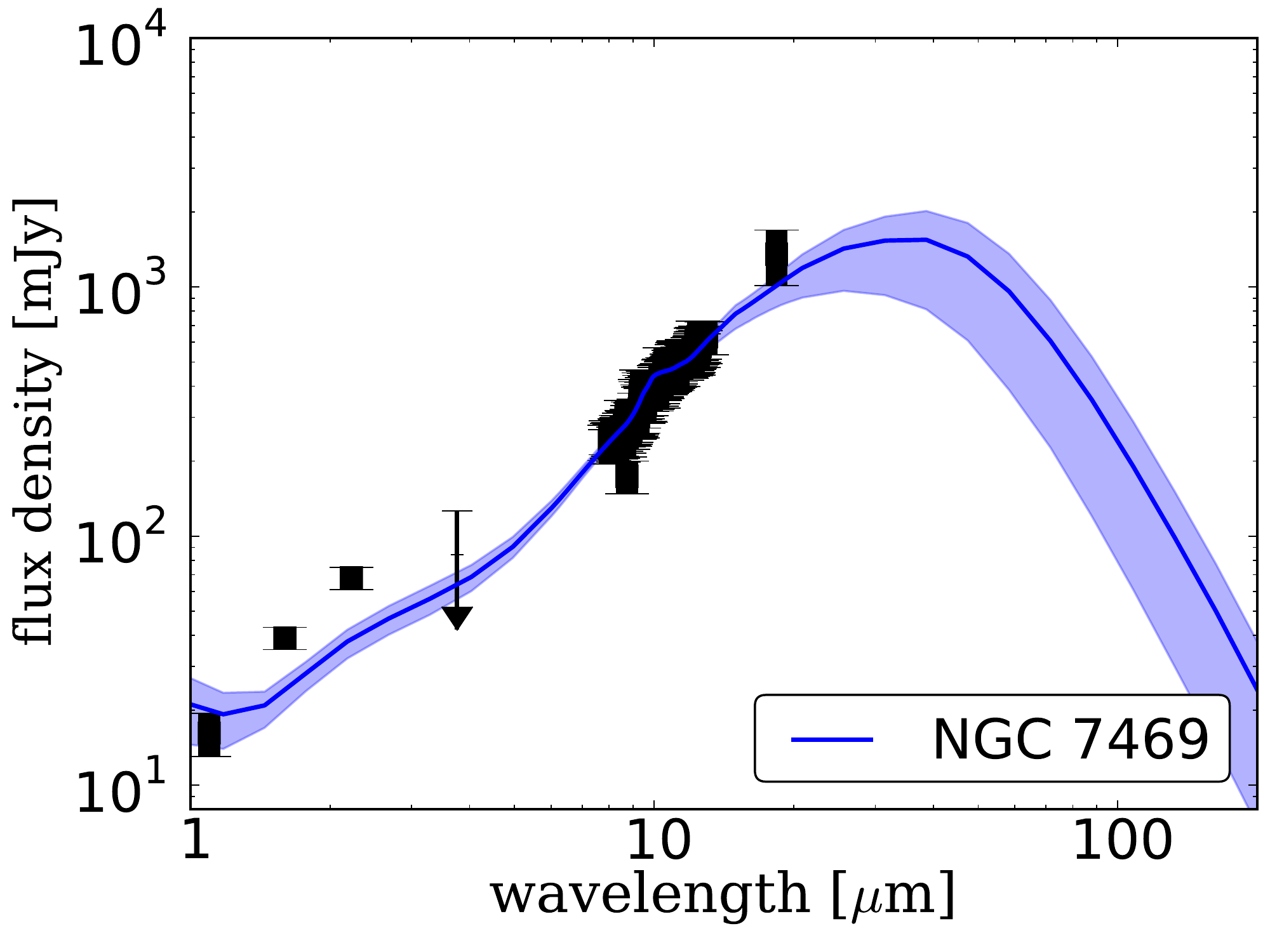}\\
{\bf \hrulefill}
HBLR
{\bf \hrulefill}
\includegraphics[width=4.6cm]{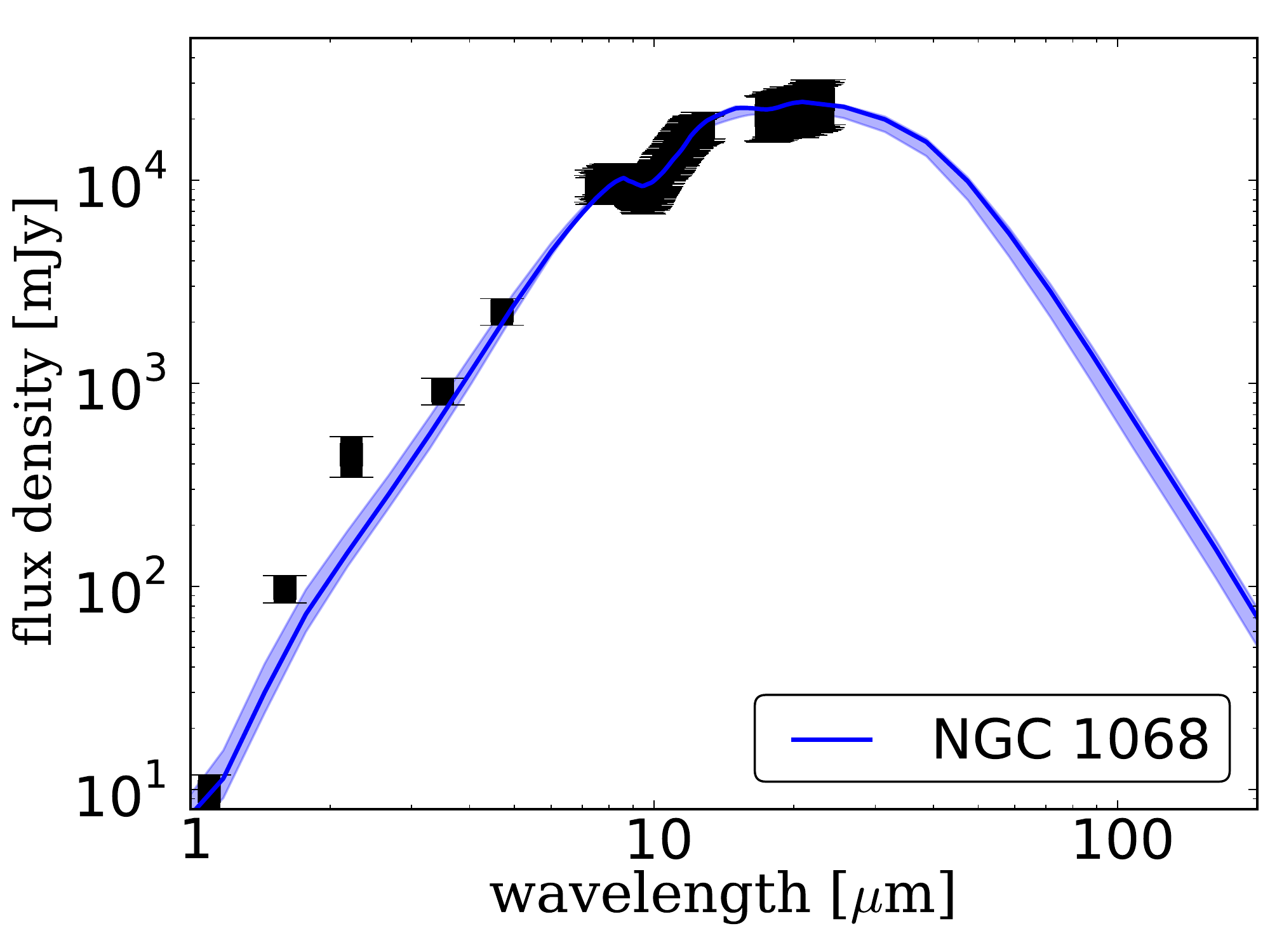}~
\includegraphics[width=4.6cm]{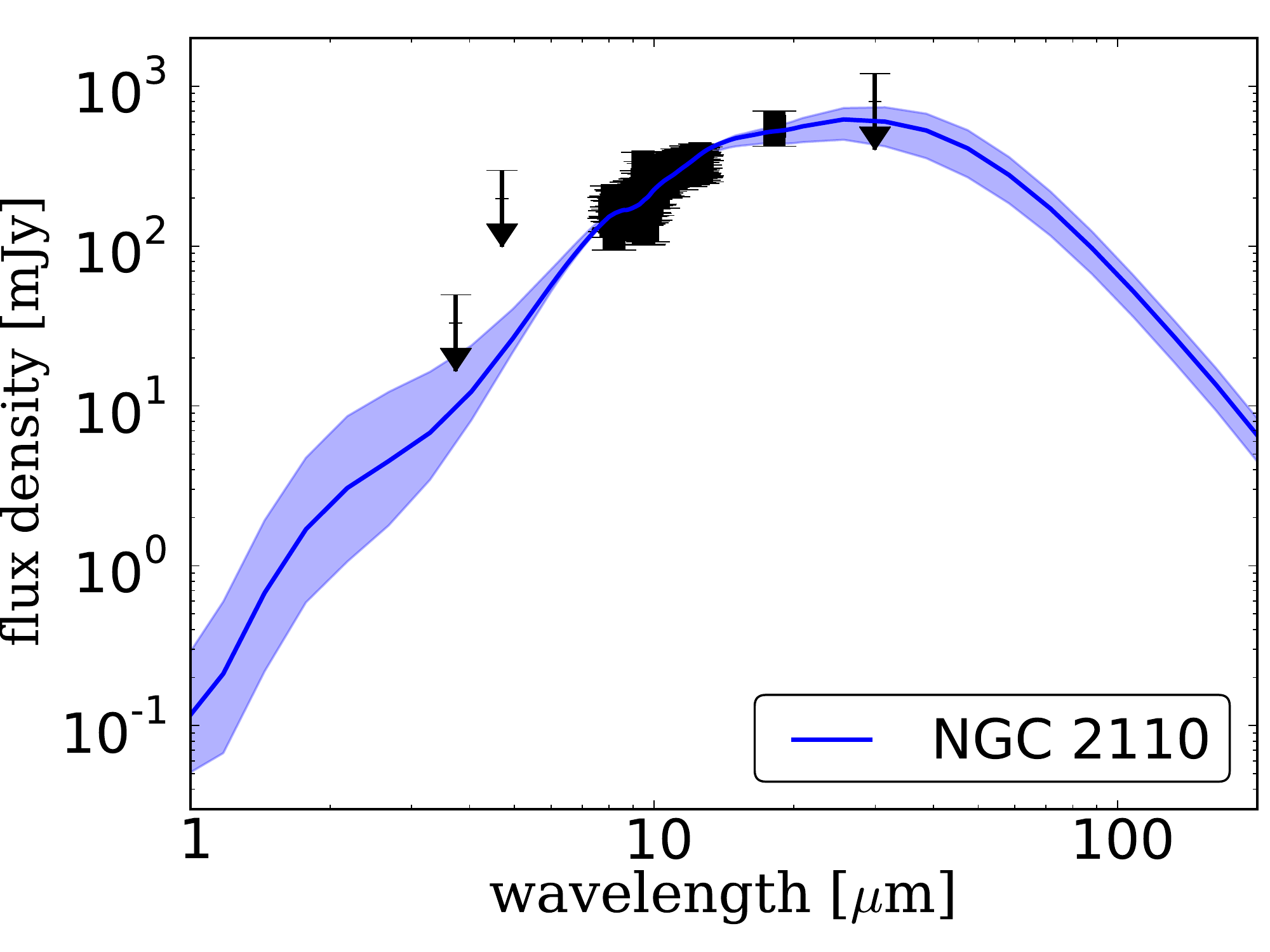}~
\includegraphics[width=4.6cm]{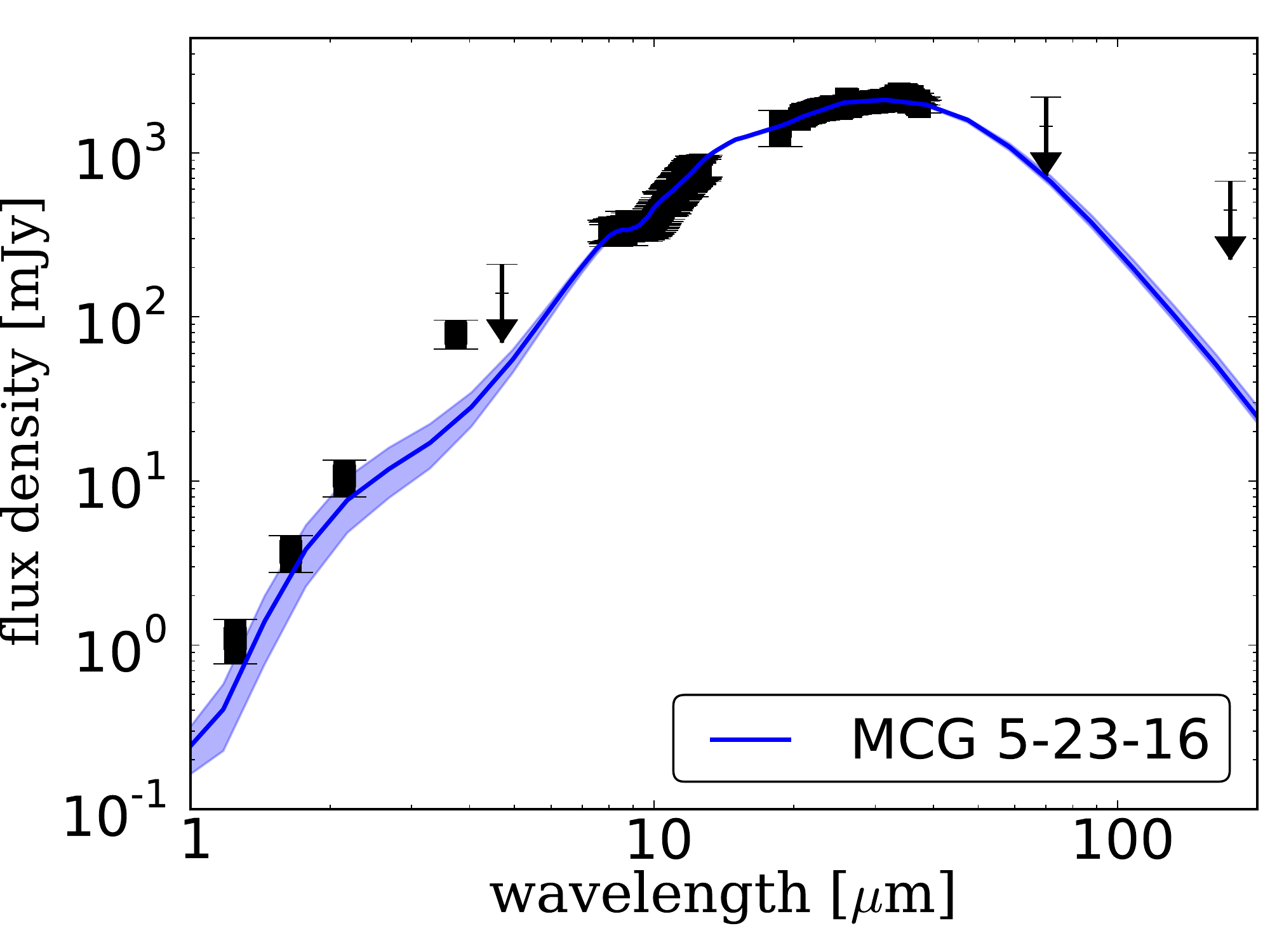}~
\includegraphics[width=4.6cm]{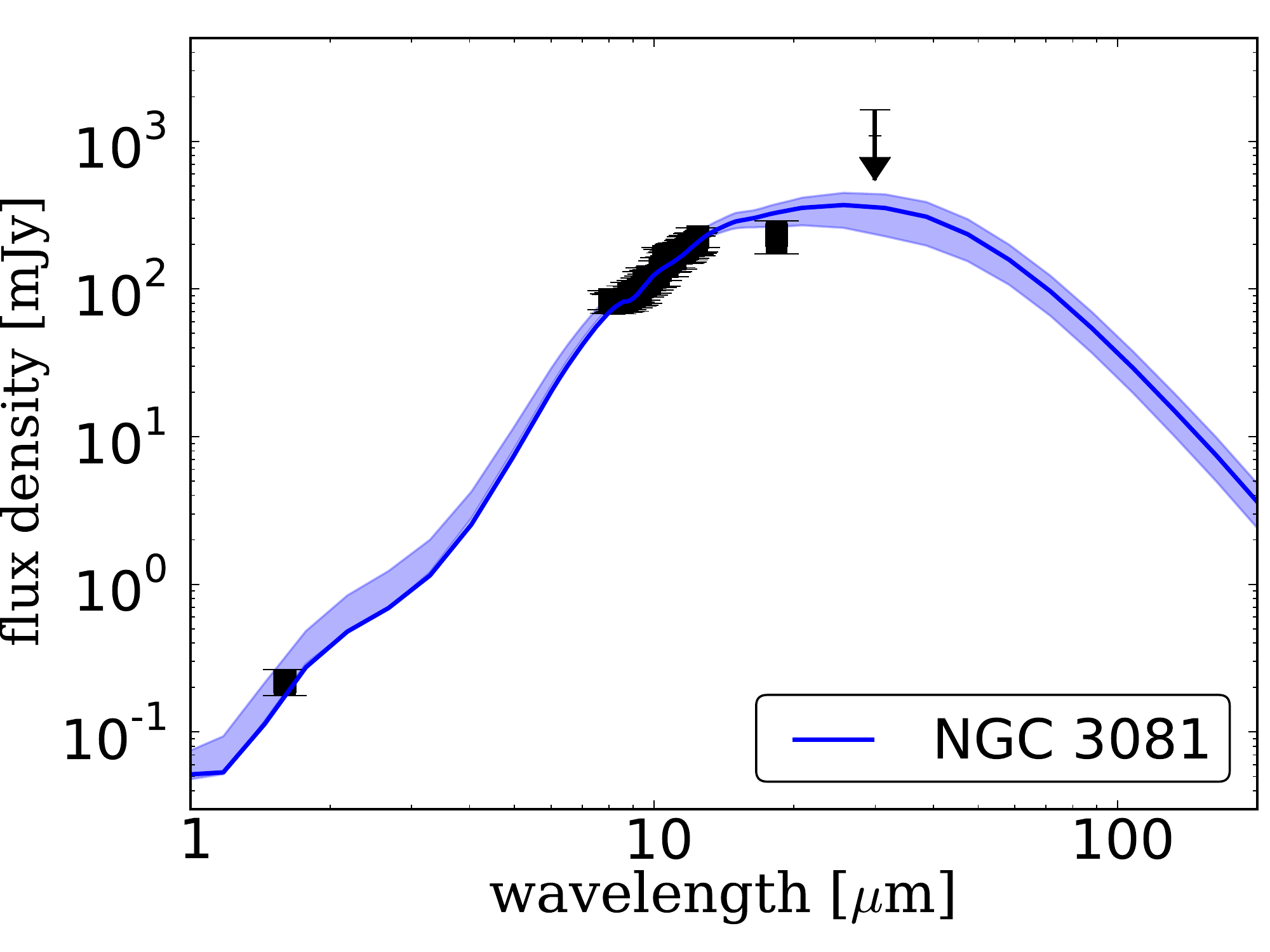}
\includegraphics[width=4.6cm]{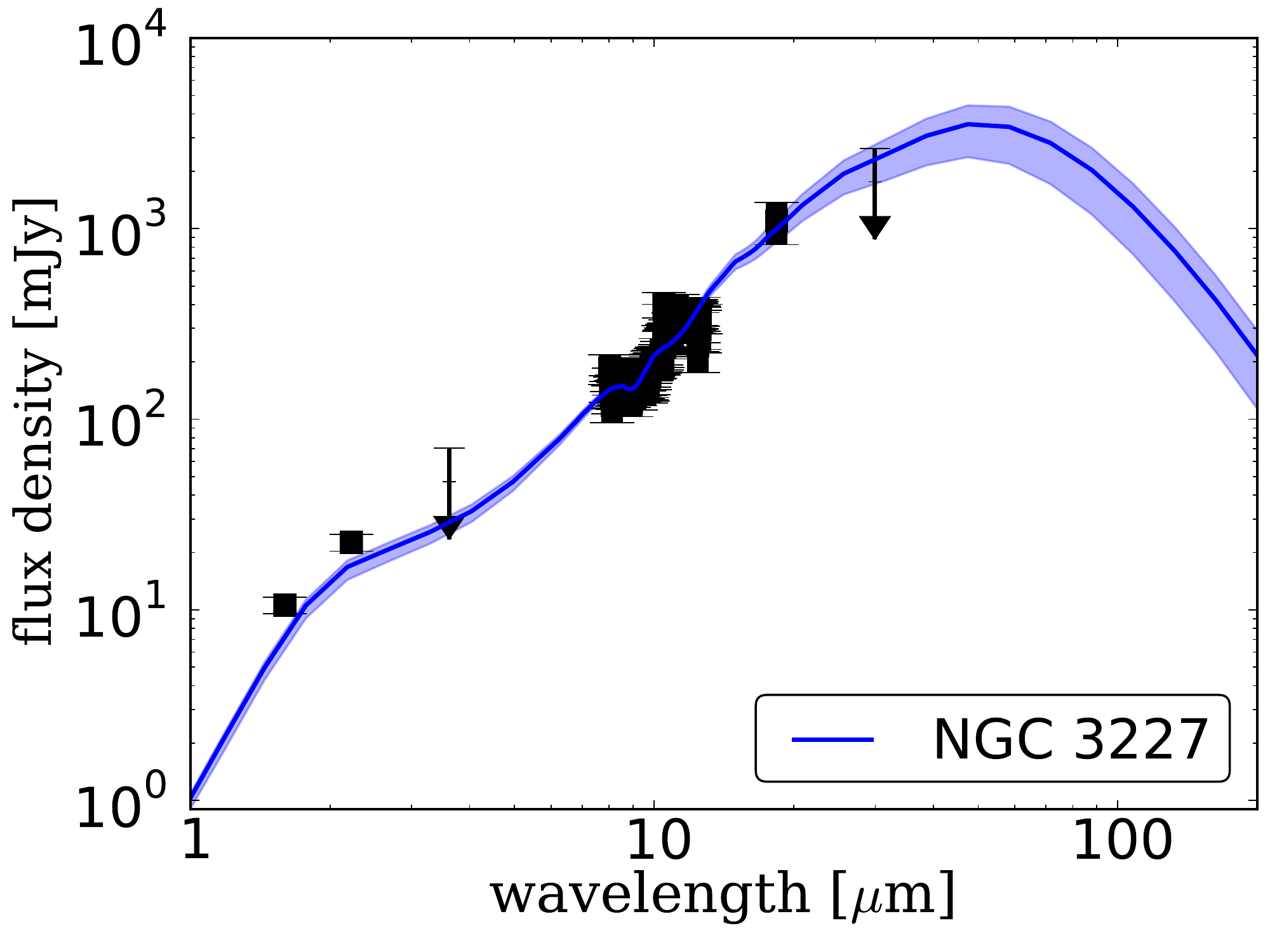}~
\includegraphics[width=4.6cm]{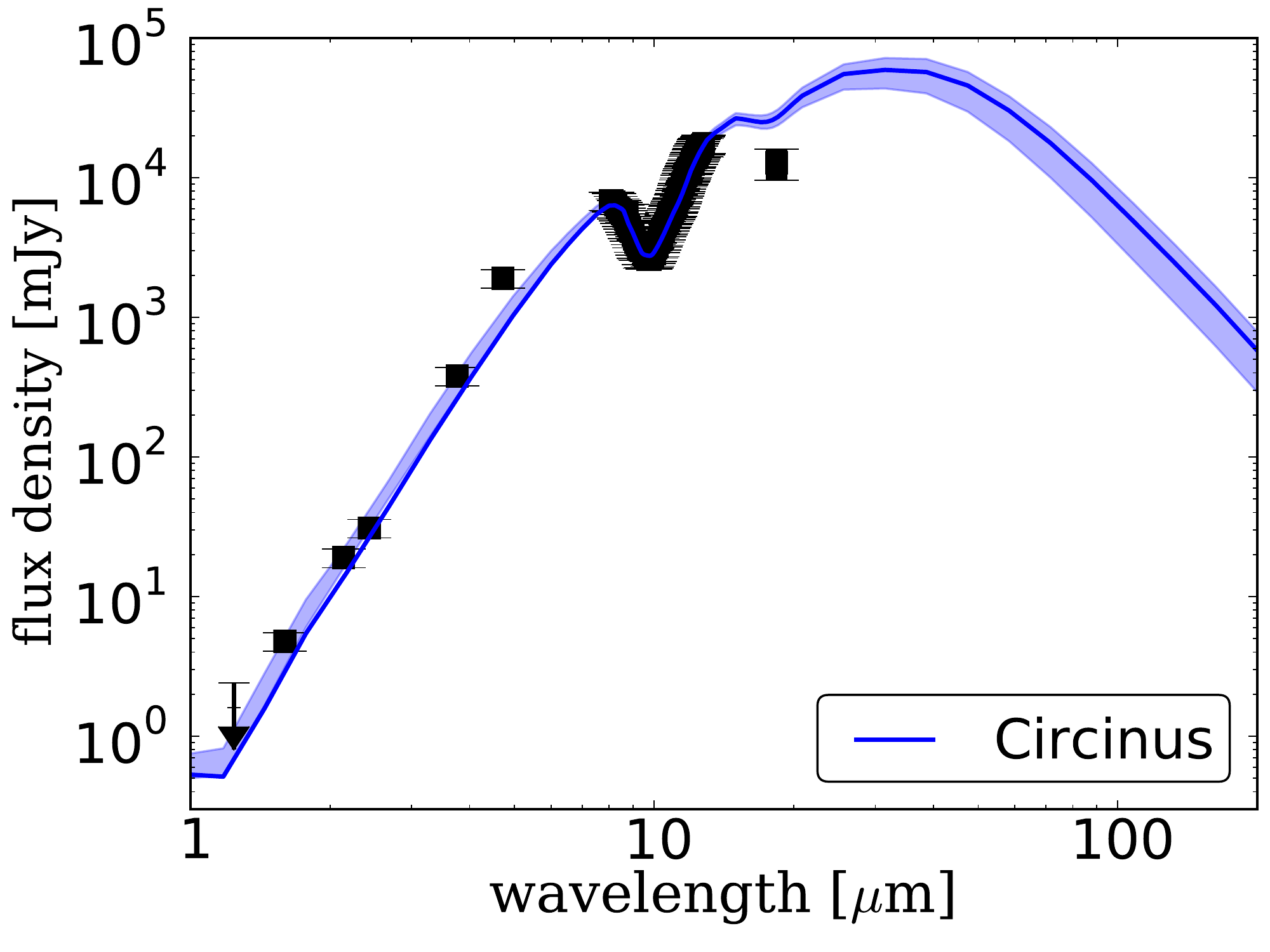}~
\includegraphics[width=4.6cm]{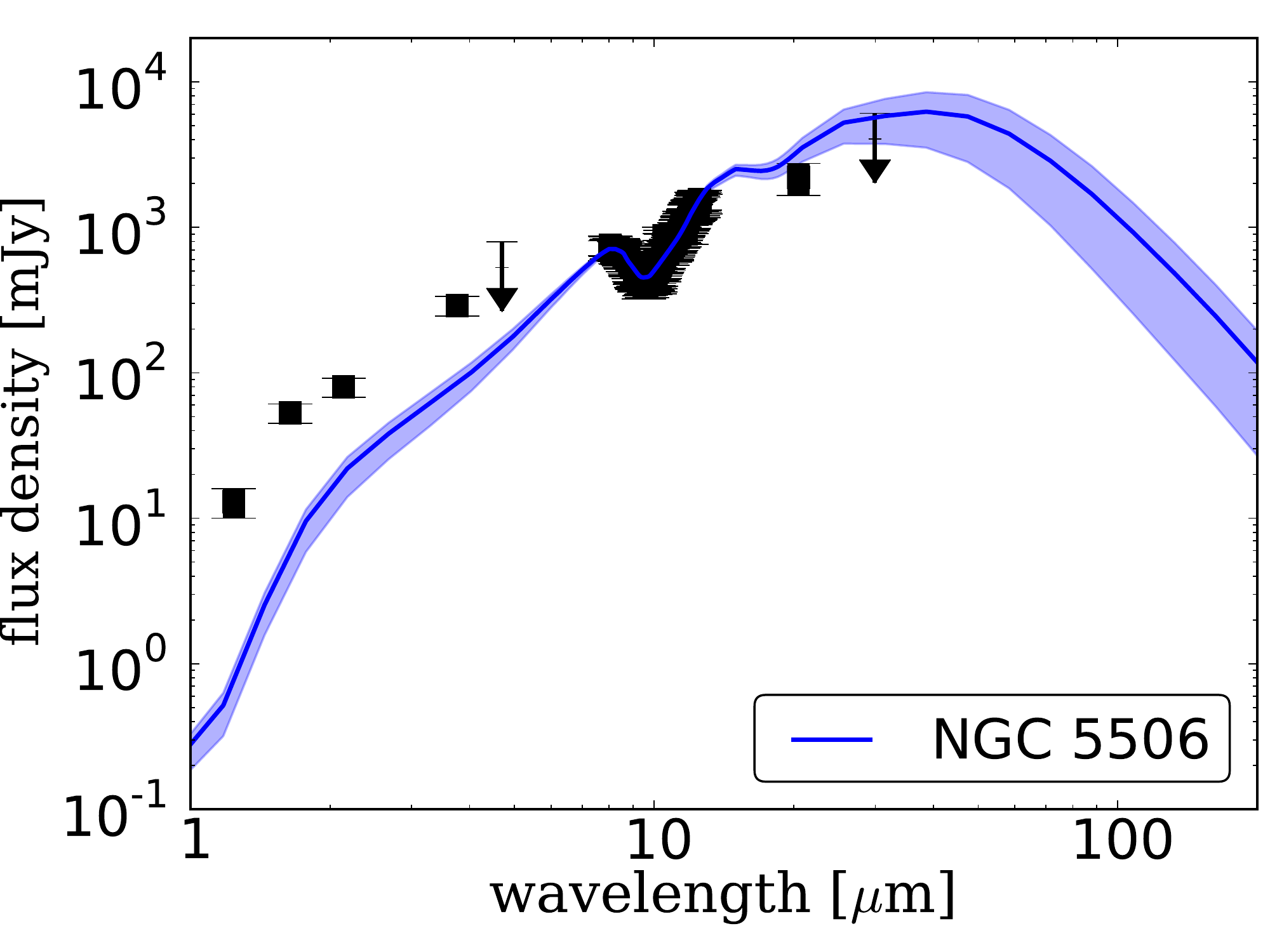}~
\includegraphics[width=4.6cm]{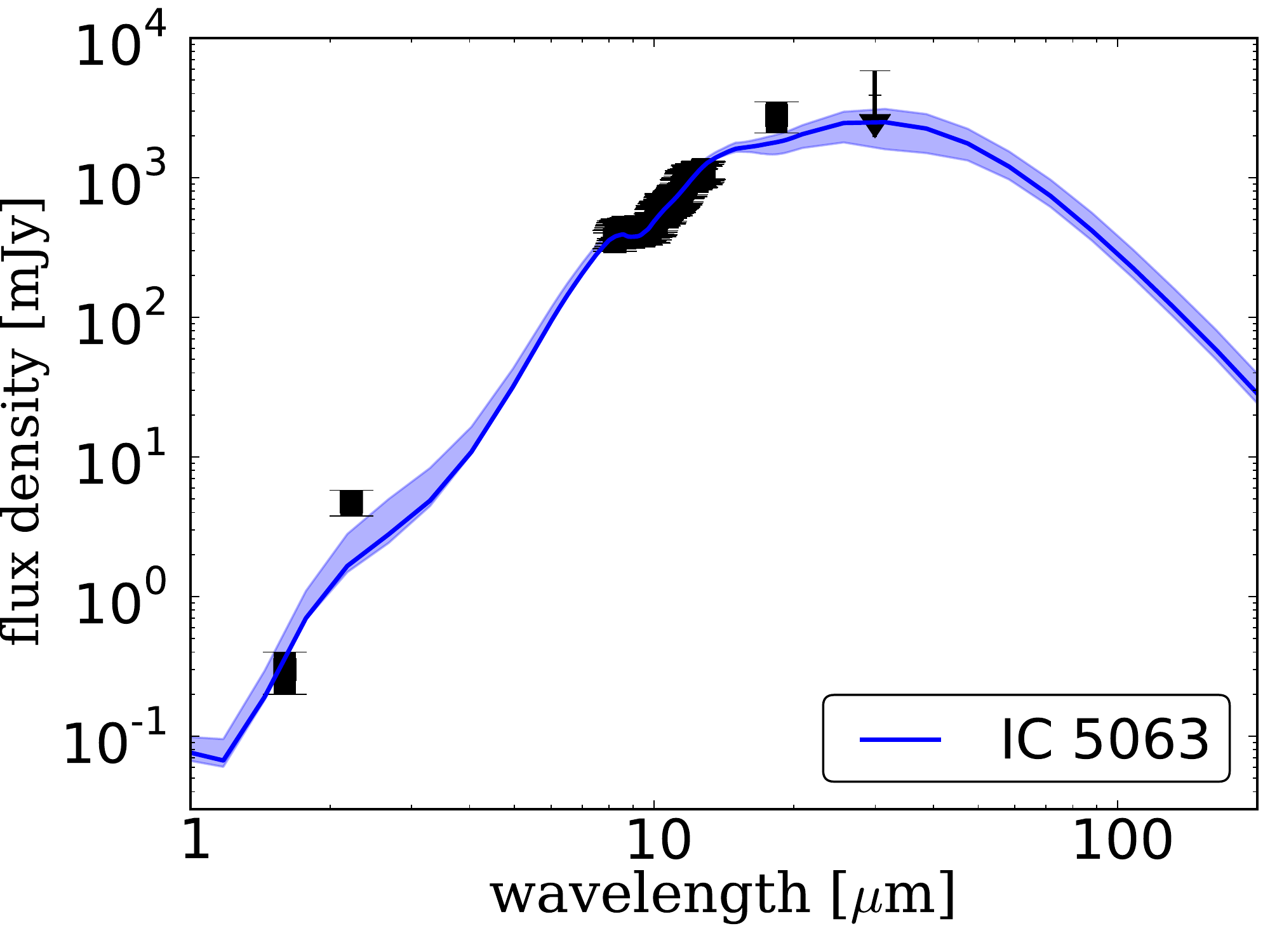}
\includegraphics[width=4.6cm]{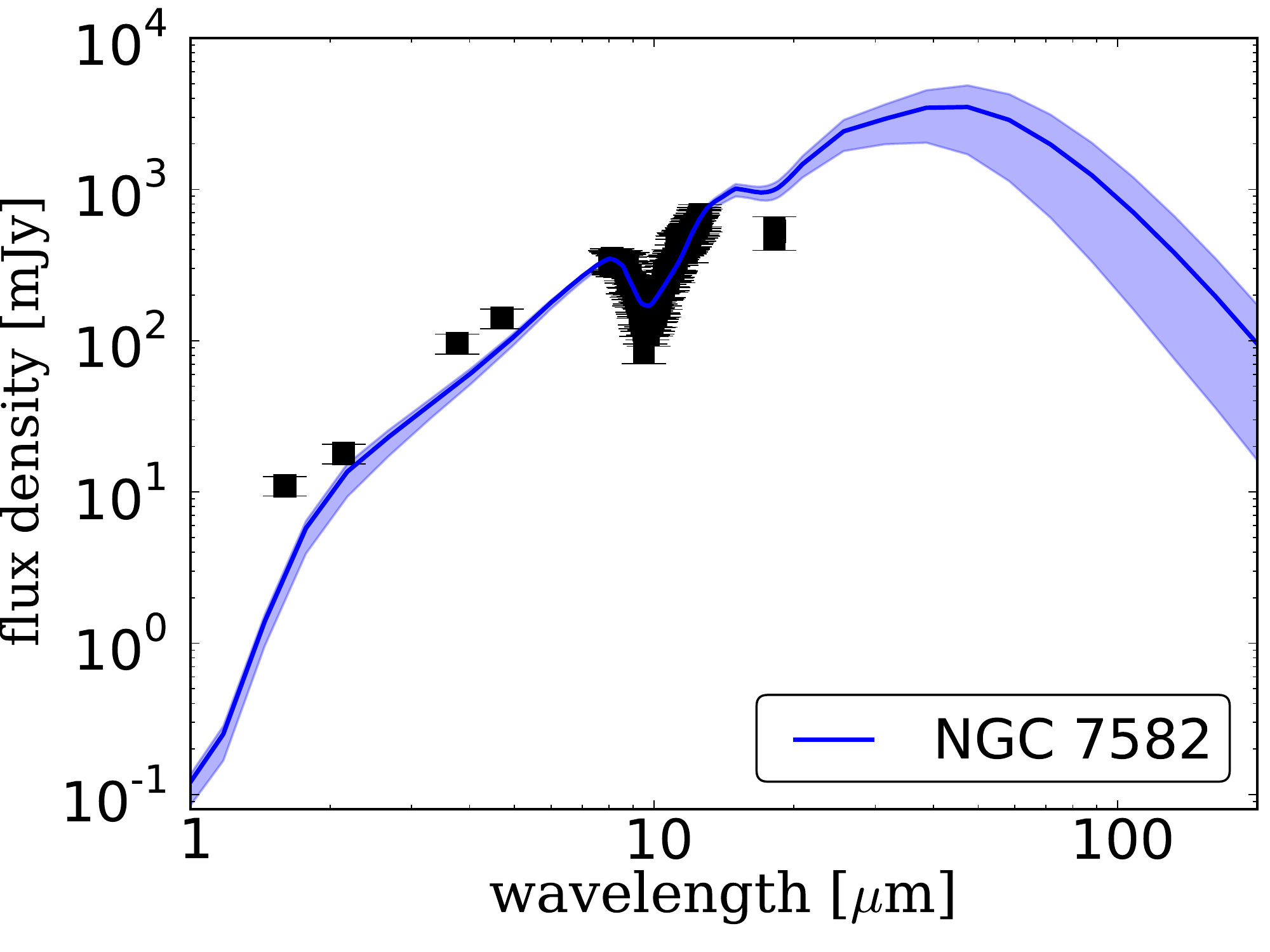}~
\includegraphics[width=4.6cm]{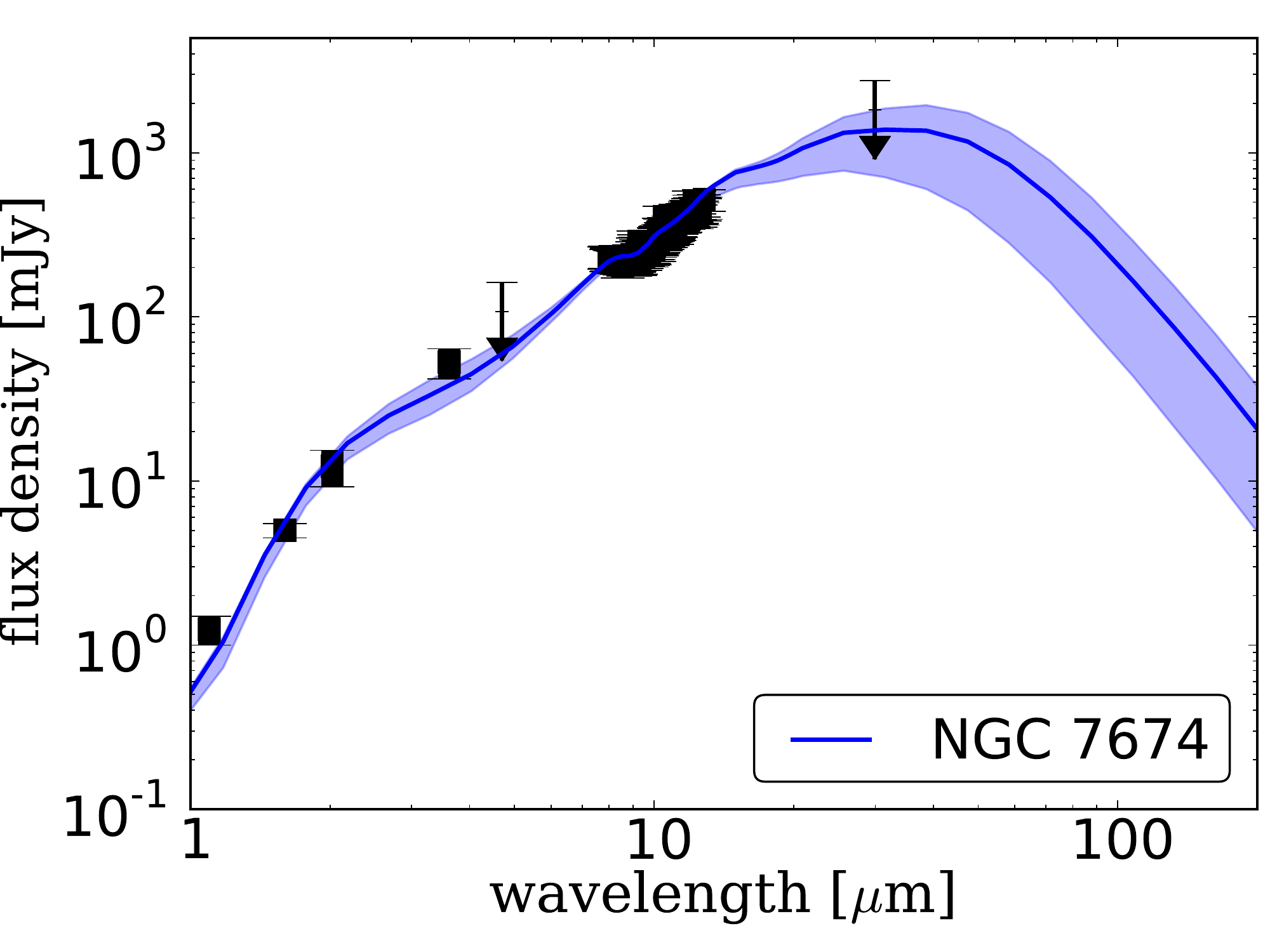}\\
{\bf \hrulefill}
NHBLR
{\bf \hrulefill}
\includegraphics[width=4.6cm]{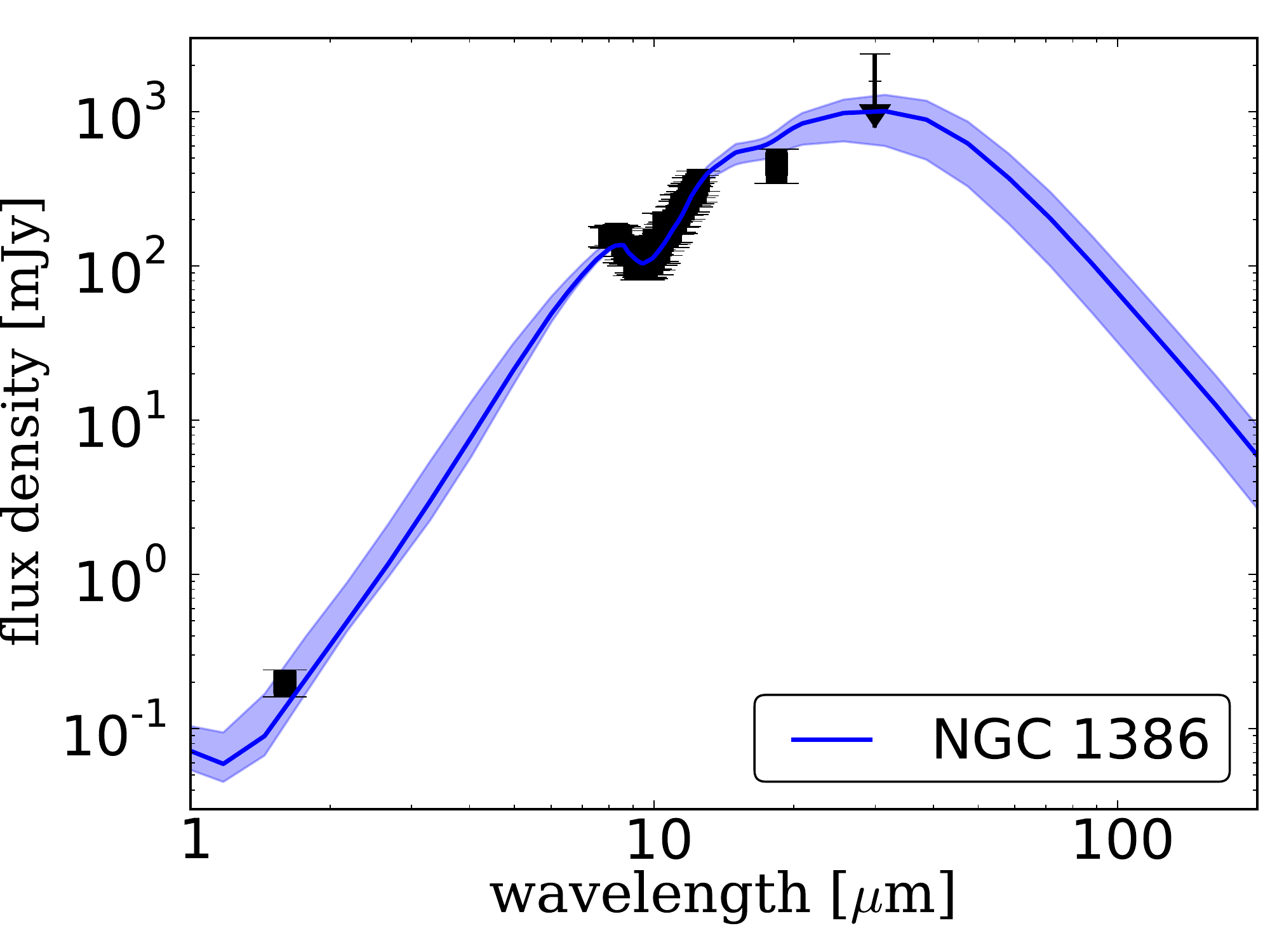}~
\includegraphics[width=4.6cm]{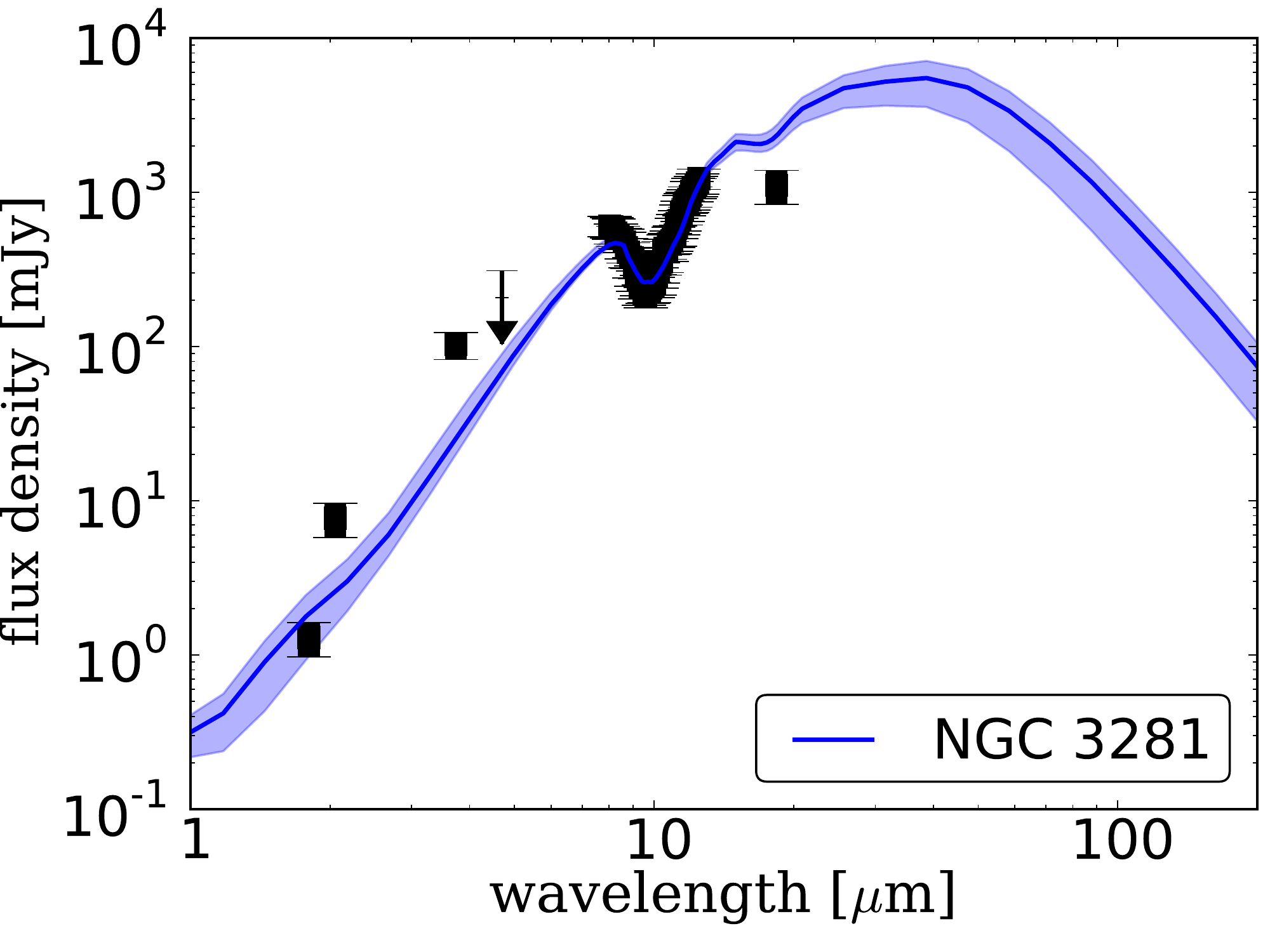}~
\includegraphics[width=4.6cm]{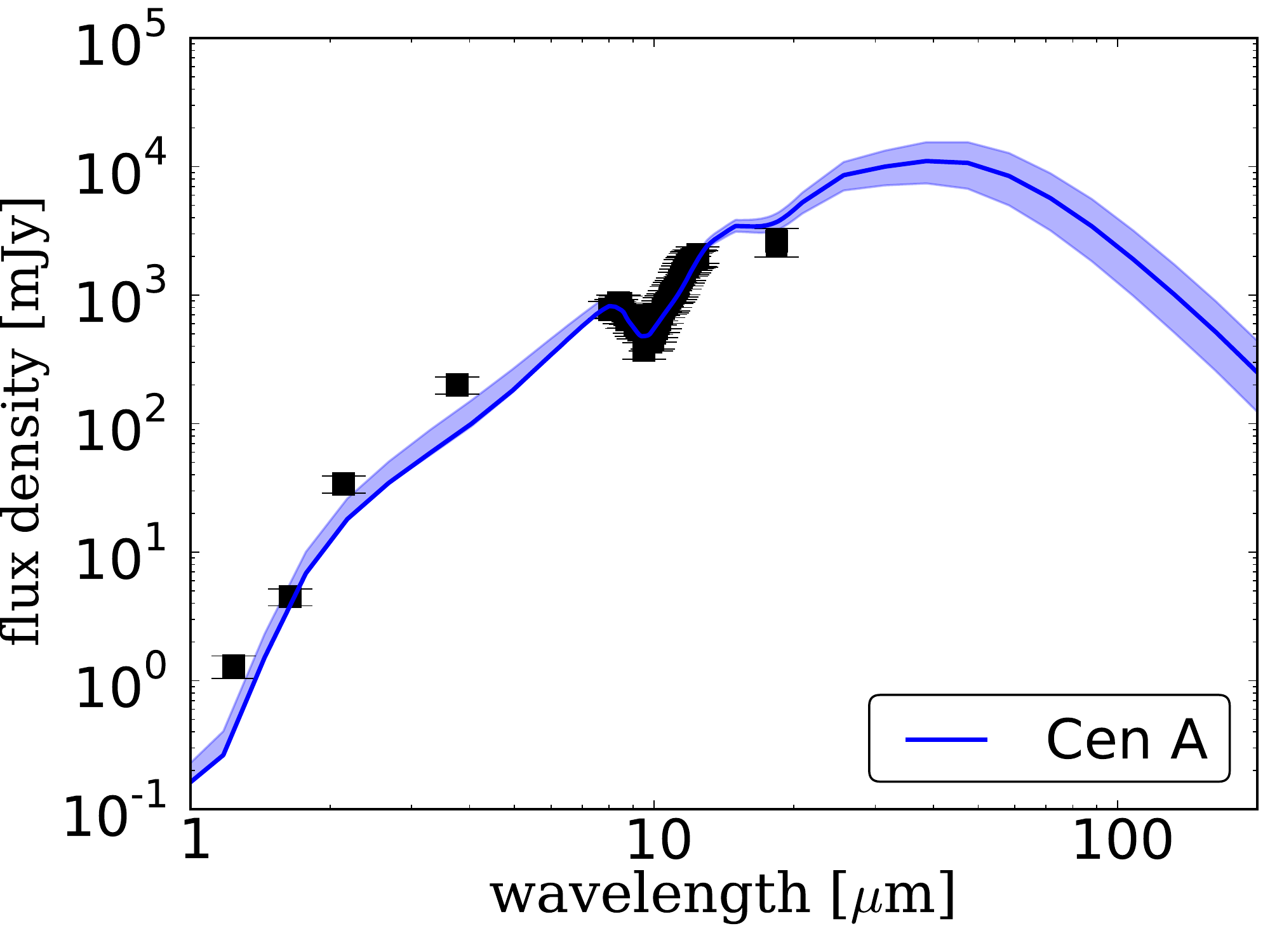}~
\includegraphics[width=4.6cm]{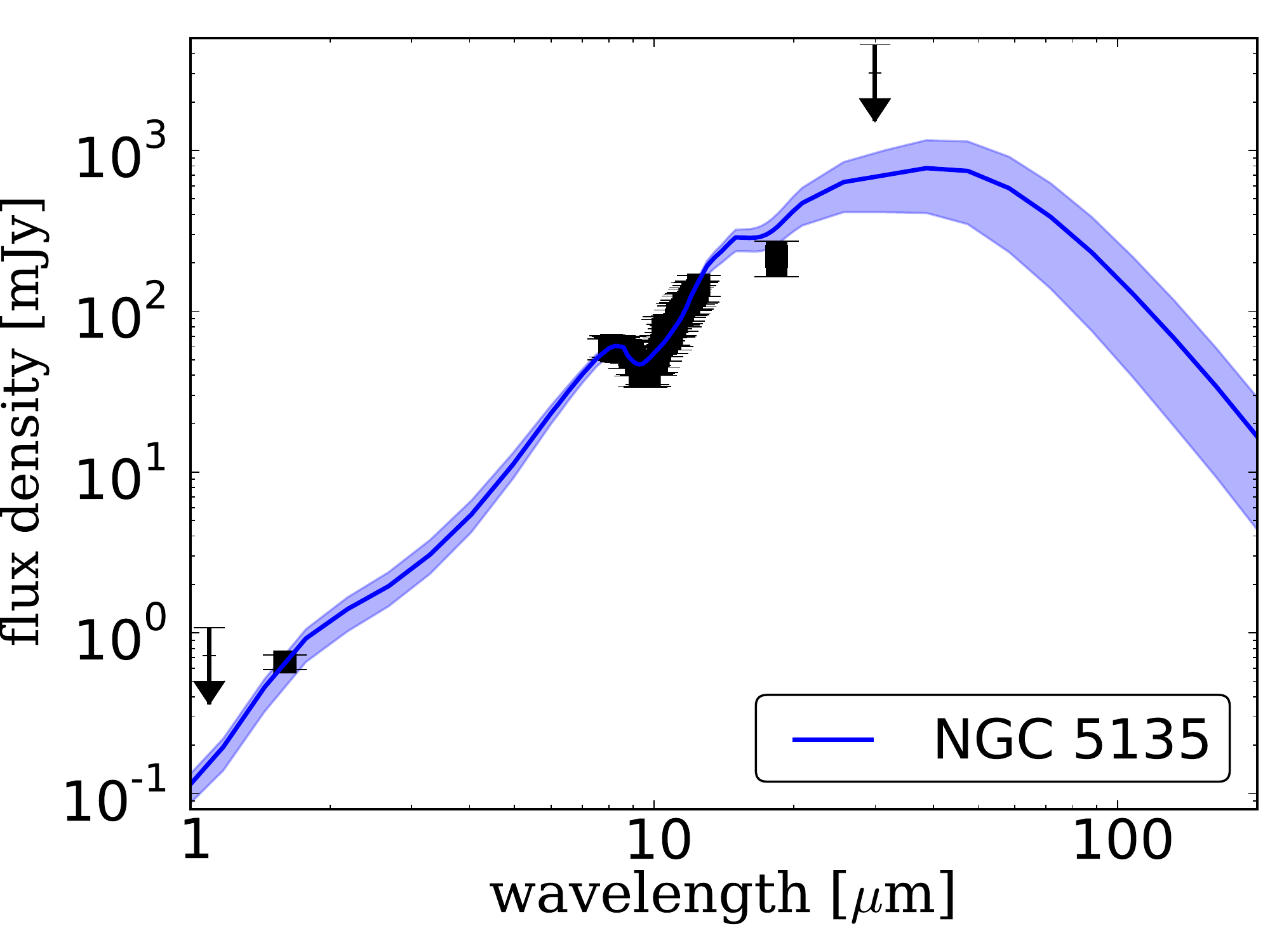}
\includegraphics[width=4.6cm]{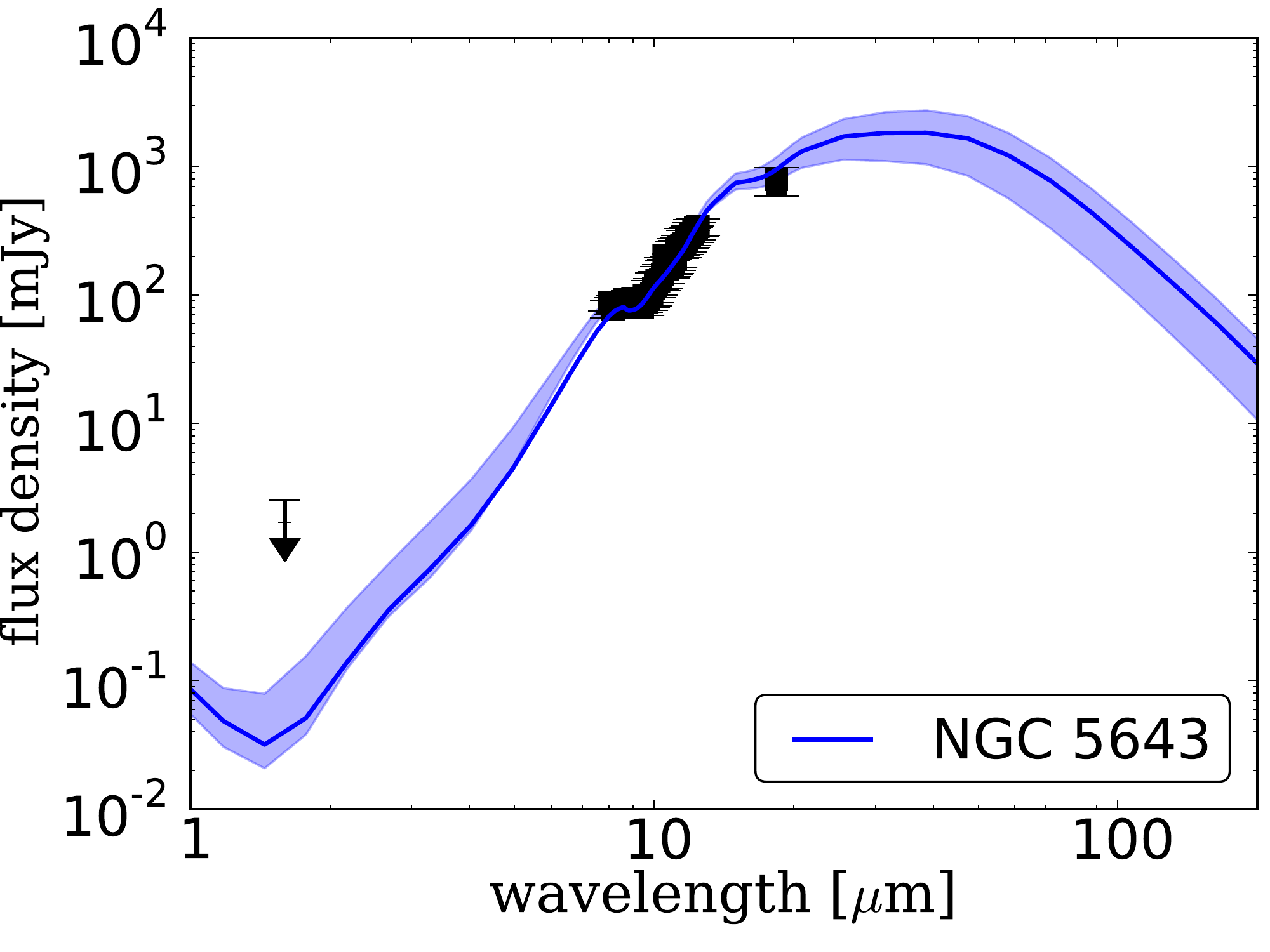}~
\includegraphics[width=4.6cm]{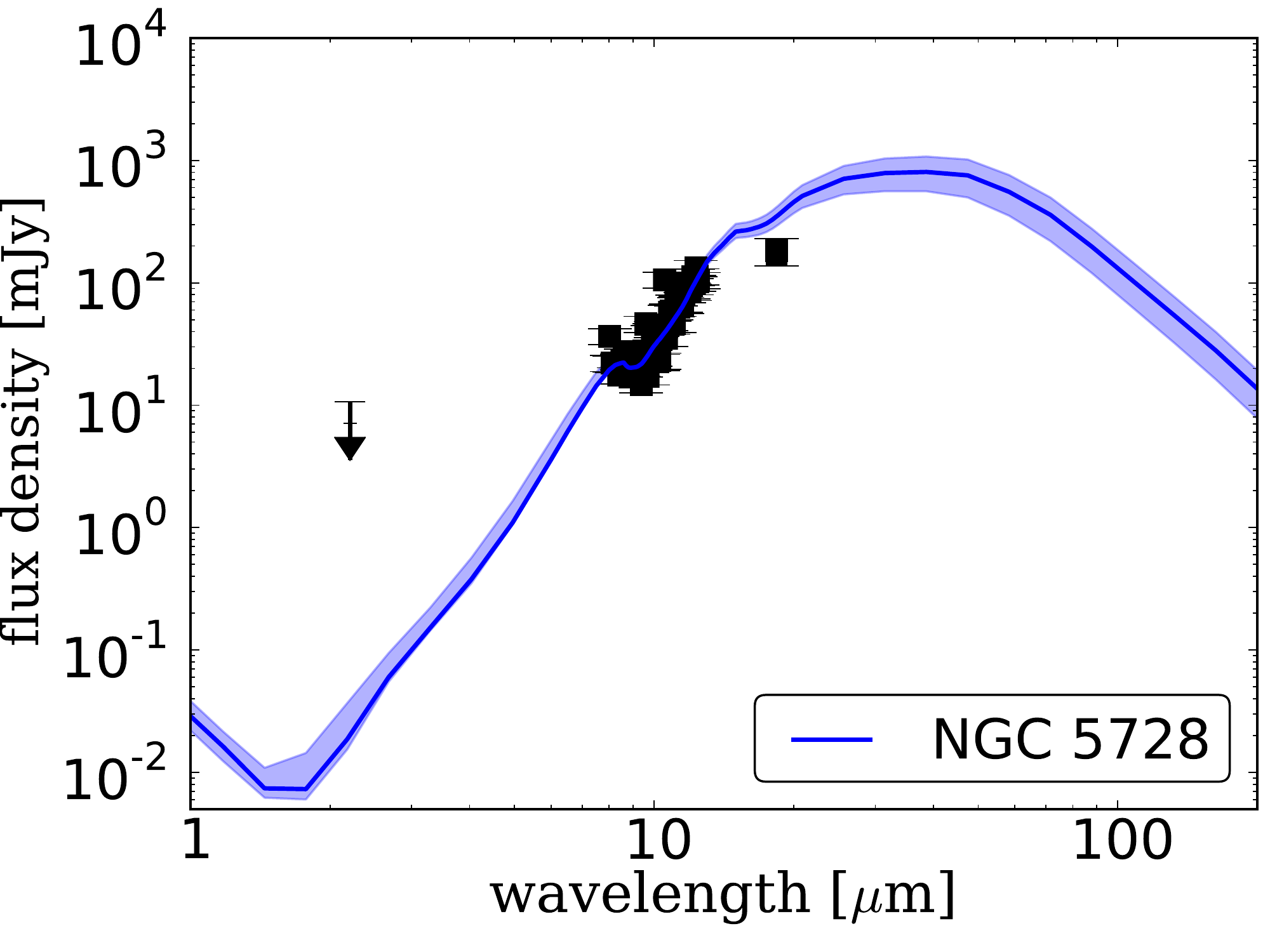}~
\includegraphics[width=4.6cm]{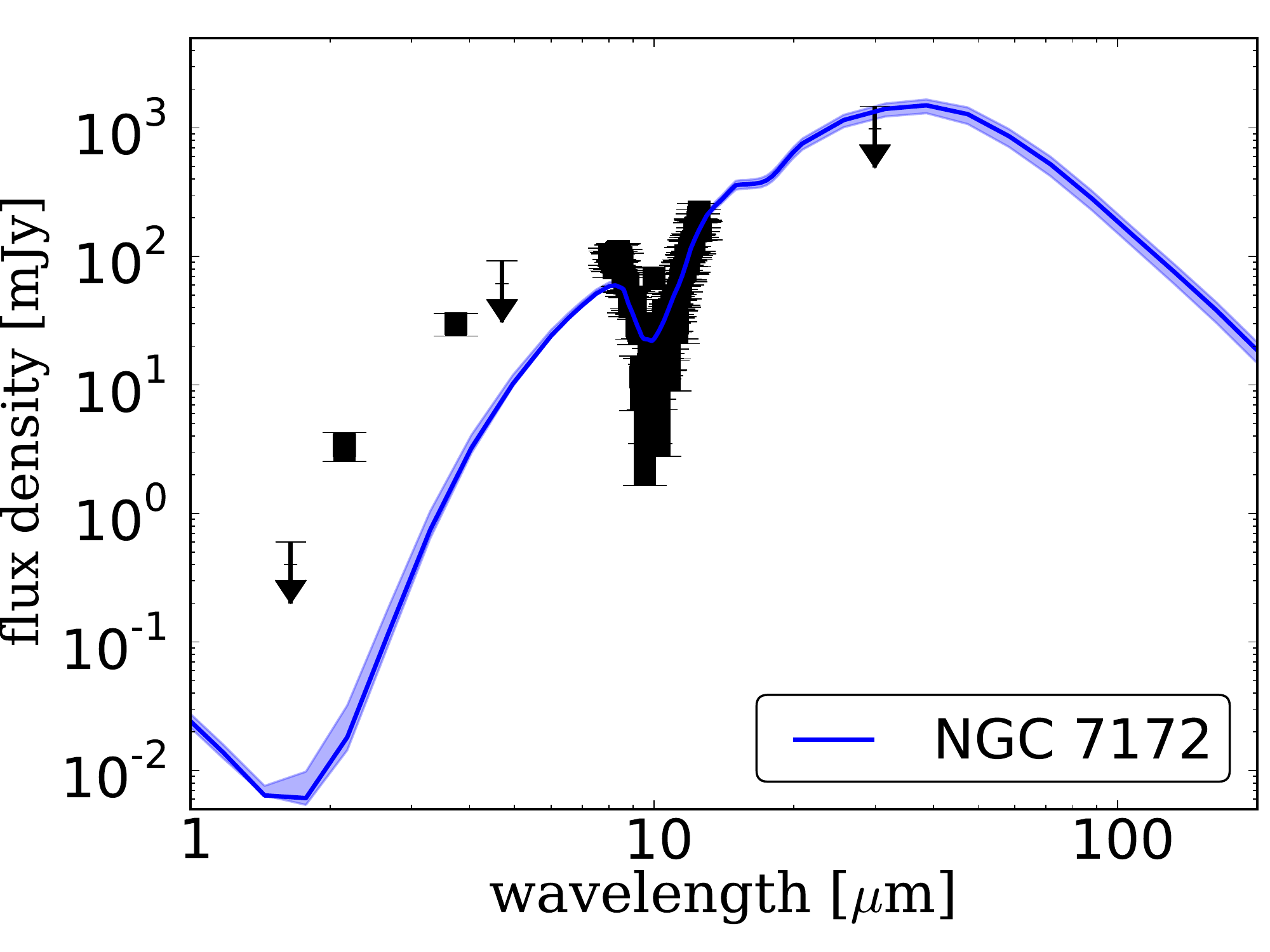}~
\caption{Clumpy torus model fits.
The filled dots are the photometric data and the black line is the MIR spectrum. The upper limit points
are shown as arrows.
The solid blue lines are the models computed with the median value of the probability distribution
of each parameter. The blue shaded areas indicate the range of models compatible with a 68\% confidence interval.
For the details on the calculation of the median values, see Section 3.}
\end{center}
\end{figure*}

\subsection{\B\ and Modeling Details}
The \textit{CLUMPY} database currently contains more than $5 \times 10^6$ models.
Therefore, when fitting the models to the observations, inherent degeneracies have to be taken into account. 
We then use the \B\ fitting tool \citep{ase09}, as it does a fast synthesis of the \textit{CLUMPY} SEDs. 
In the last version of \B\, the inference over the model parameters can be done either using neural network 
interpolation or multilinear interpolation in the full database. After running several tests, \cite{ram14}
concluded that the latter interpolation produces more robust results. Therefore, here we use linear
interpolation, which results in slight differences in the
fitted parameters (within 1$\sigma$ for the majority of the fits)
for the 13 galaxies that were modeled by \cite{alo11} using the neural network interpolation and subsequently re-fitted in this paper.

\B\ can be used to fit photometry and/or spectra in a Bayesian scheme, carrying out inference over the model 
parameters for observed SEDs. This way we can specify a-priori information about the model parameters. 
Here we consider uniform prior distribution in the range of each parameter, as summarized in Table~3. 
The prior distribution of inclination angle ($i$) is fixed from previous observations if available in the literature, 
following the same approach as in \cite{alo11}. 
From the objects in our sample taken from \cite{gon13}, 
NGC~1386 has two possible inclination angles 65$^{\circ}$ and 85$^{\circ}$ \citep{lev06}.
Thus, we use a uniform prior in the range 60$^{\circ}$--90$^{\circ}$ for this source.
For the galaxies taken from \cite{alo11}, we use the same inclination angle constraints they employed, which are compiled in 
column 11 in Table~1.

We finally include the AGN direct emission (i.e. a broken power-law) 
which is defined in Eq.~(13) of \cite{nen08a} to the SED for type-1 AGN, 
in order to reproduce the flatter  slope of the NIR band \citep[see][ for further details]{ram14}.

\begin{table*}
\begin{center}
\scriptsize
\caption{Fitted torus  model parameters from SED + spectroscopy data}
\begin{tabular}{lccccccccccc}
\hline
\hline
Galaxy & 
\multicolumn{1}{c}{$\sigma_{\rm torus}$} &
\multicolumn{1}{c}{$Y$} &
\multicolumn{1}{c}{$N_0$} &
\multicolumn{1}{c}{$q$} &
\multicolumn{1}{c}{$\tau_{\rm V}$} &
\multicolumn{1}{c}{$i$} & 
\multicolumn{1}{c}{$C_{\rm T}$}  &
\multicolumn{1}{c}{$\log L_{\rm bol}^{(\rm mod)}$}   &
\multicolumn{1}{c}{$r_{\rm in}$} &
\multicolumn{1}{c}{$r_{\rm out}$} &
\multicolumn{1}{c}{$H$} 
\\

& [deg]  & & & & & [deg] & & [erg/s]  & [pc] & [pc] & [pc] \\
\hline
\multicolumn{12}{c}{Type-1}\\
\hline
NGC~1365 & $ 19_{ -2}^{+ 5}$ & $ 24_{ -4}^{+ 3}$ & $ 9_{ -3}^{+ 2}$ &$ 0.3_{-0.2}^{+ 0.3}$ & $ 79_{ -35}^{+ 42}$ & $ 19_{ -11}^{+ 19}$ & $ 0.16_{- 0.04}^{+ 0.08}$ & $43.2_{- 0.1}^{+ 0.8}$ & $0.053_{-0.005}^{+0.005}$ & $ 1.3_{- 0.3}^{+ 0.2}$ & $ 0.4_{- 0.1}^{+ 0.1}$\\
NGC~4151 & $ 16_{ -1}^{+ 1}$ & $ 19_{ -4}^{+ 4}$ & $ 13_{ -1}^{+ 1}$ &$ 1.6_{-0.3}^{+ 0.3}$ & $ 89_{ -8}^{+ 8}$ & $ 71_{ -2}^{+ 2}$ & $ 0.13_{- 0.02}^{+ 0.03}$ & $43.9_{- 0.1}^{+ 0.9}$ & $0.108_{-0.008}^{+0.007}$ & $ 2.1_{- 0.5}^{+ 0.5}$ & $ 0.6_{- 0.1}^{+ 0.2}$\\
IC~4329A & $ 40_{ -1}^{+ 1}$ & $ 8_{ -1}^{+ 1}$ & $ 12_{ -1}^{+ 1}$ &$ 0.5_{-0.1}^{+ 0.1}$ & $ 148_{ -2}^{+ 1}$ & $ 4_{ -3}^{+ 4}$ & $ 0.58_{- 0.04}^{+ 0.04}$ & $44.4_{- 0.1}^{+ 1.8}$ & $0.192_{-0.001}^{+0.002}$ & $ 1.8_{- 0.1}^{+ 0.1}$ & $ 1.1_{- 0.0}^{+ 0.0}$\\
NGC~7469 & $ 21_{ -2}^{+ 2}$ & $ 22_{ -4}^{+ 4}$ & $ 13_{ -1}^{+ 1}$ &$ 1.3_{-0.3}^{+ 0.3}$ & $ 124_{ -14}^{+ 12}$ & $ 59_{ -4}^{+ 3}$ & $ 0.20_{- 0.04}^{+ 0.05}$ & $44.6_{- 0.1}^{+ 0.8}$ & $0.239_{-0.021}^{+0.020}$ & $ 5.3_{- 1.2}^{+ 1.1}$ & $ 1.9_{- 0.4}^{+ 0.4}$\\
\hline
\multicolumn{12}{c}{HBLR}\\
\hline
NGC~1068 & $ 56_{ -18}^{+ 8}$ & $ 6_{ -1}^{+ 2}$ & $ 5_{ -1}^{+ 3}$ &$ 0.6_{-0.4}^{+ 1.2}$ & $ 38_{ -7}^{+ 3}$ & $ 67_{ -5}^{+ 11}$ & $ 0.78_{- 0.19}^{+ 0.06}$ & $44.4_{- 0.1}^{+ 0.9}$ & $0.198_{-0.006}^{+0.015}$ & $ 1.3_{- 0.1}^{+ 0.6}$ & $ 1.1_{- 0.1}^{+ 0.3}$\\
NGC~2110 & $ 55_{ -8}^{+ 8}$ & $ 17_{ -6}^{+ 7}$ & $ 9_{ -2}^{+ 2}$ &$ 2.7_{-0.3}^{+ 0.2}$ & $ 146_{ -4}^{+ 2}$ & $ 40_{ -6}^{+ 5}$ & $ 0.88_{- 0.11}^{+ 0.05}$ & $43.3_{- 0.1}^{+ 1.0}$ & $0.058_{-0.002}^{+0.003}$ & $ 1.0_{- 0.4}^{+ 0.5}$ & $ 0.8_{- 0.4}^{+ 0.4}$\\
MCG~-5-23-16 & $ 58_{ -8}^{+ 5}$ & $ 20_{ -1}^{+ 2}$ & $ 7_{ -1}^{+ 2}$ &$ 2.1_{-0.1}^{+ 0.2}$ & $ 144_{ -6}^{+ 3}$ & $ 48_{ -3}^{+ 7}$ & $ 0.85_{- 0.05}^{+ 0.03}$ & $43.9_{- 0.1}^{+ 1.2}$ & $0.114_{-0.003}^{+0.004}$ & $ 2.3_{- 0.2}^{+ 0.3}$ & $ 2.0_{- 0.3}^{+ 0.3}$\\
NGC~3081 & $ 62_{ -5}^{+ 4}$ & $ 11_{ -4}^{+ 9}$ & $ 12_{ -1}^{+ 1}$ &$ 2.6_{-0.4}^{+ 0.2}$ & $ 98_{ -11}^{+ 12}$ & $ 66_{ -19}^{+ 13}$ & $ 0.96_{- 0.03}^{+ 0.02}$ & $43.1_{- 0.1}^{+ 1.0}$ & $0.043_{-0.003}^{+0.003}$ & $ 0.4_{- 0.1}^{+ 0.5}$ & $ 0.3_{- 0.1}^{+ 0.4}$\\
NGC~3227 & $ 57_{ -1}^{+ 1}$ & $ 20_{ -1}^{+ 1}$ & $ 13_{ -1}^{+ 1}$ &$ 0.0_{-0.0}^{+ 0.1}$ & $ 147_{ -2}^{+ 1}$ & $ 6_{ -4}^{+ 5}$ & $ 0.94_{- 0.02}^{+ 0.01}$ & $43.0_{- 0.1}^{+ 1.0}$ & $0.041_{-0.003}^{+0.003}$ & $ 0.8_{- 0.1}^{+ 0.1}$ & $ 0.7_{- 0.1}^{+ 0.1}$\\
Circinus & $ 64_{ -5}^{+ 3}$ & $ 16_{ -1}^{+ 2}$ & $ 12_{ -2}^{+ 1}$ &$ 0.4_{-0.2}^{+ 0.3}$ & $ 35_{ -2}^{+ 2}$ & $ 37_{ -7}^{+ 8}$ & $ 0.96_{- 0.03}^{+ 0.02}$ & $43.4_{- 0.1}^{+ 1.0}$ & $0.062_{-0.003}^{+0.004}$ & $ 1.0_{- 0.2}^{+ 0.2}$ & $ 0.9_{- 0.1}^{+ 0.2}$\\
NGC~5506 & $ 48_{ -3}^{+ 6}$ & $ 16_{ -2}^{+ 3}$ & $ 10_{ -3}^{+ 2}$ &$ 0.2_{-0.2}^{+ 0.3}$ & $ 79_{ -5}^{+ 4}$ & $ 32_{ -1}^{+ 4}$ & $ 0.79_{- 0.05}^{+ 0.04}$ & $44.0_{- 0.1}^{+ 1.0}$ & $0.130_{-0.007}^{+0.008}$ & $ 2.1_{- 0.5}^{+ 0.5}$ & $ 1.6_{- 0.3}^{+ 0.4}$\\
IC~5063 & $ 61_{ -6}^{+ 4}$ & $ 14_{ -7}^{+ 7}$ & $ 12_{ -1}^{+ 1}$ &$ 2.5_{-1.1}^{+ 0.2}$ & $ 101_{ -9}^{+ 7}$ & $ 77_{ -12}^{+ 7}$ & $ 0.96_{- 0.04}^{+ 0.02}$ & $44.3_{- 0.1}^{+ 1.0}$ & $0.182_{-0.008}^{+0.009}$ & $ 2.6_{- 1.4}^{+ 1.6}$ & $ 2.3_{- 1.3}^{+ 1.4}$\\
NGC~7582 & $ 53_{ -2}^{+ 3}$ & $ 20_{ -1}^{+ 2}$ & $ 12_{ -2}^{+ 1}$ &$ 0.1_{-0.0}^{+ 0.1}$ & $ 79_{ -9}^{+ 7}$ & $ 6_{ -4}^{+ 5}$ & $ 0.90_{- 0.02}^{+ 0.02}$ & $43.5_{- 0.1}^{+ 0.9}$ & $0.070_{-0.004}^{+0.005}$ & $ 1.4_{- 0.2}^{+ 0.3}$ & $ 1.2_{- 0.2}^{+ 0.2}$\\
NGC~7674 & $ 39_{ -9}^{+ 13}$ & $ 15_{ -4}^{+ 5}$ & $ 8_{ -3}^{+ 3}$ &$ 1.1_{-0.6}^{+ 0.5}$ & $ 133_{ -15}^{+ 9}$ & $ 44_{ -13}^{+ 13}$ & $ 0.56_{- 0.20}^{+ 0.17}$ & $44.8_{- 0.1}^{+ 0.5}$ & $0.330_{-0.038}^{+0.068}$ & $ 5.3_{- 1.7}^{+ 2.5}$ & $ 3.3_{- 1.0}^{+ 1.4}$\\
\hline
\multicolumn{12}{c}{NHBLR}\\
\hline
NGC~1386 & $ 56_{ -9}^{+ 7}$ & $ 19_{ -5}^{+ 5}$ & $ 8_{ -1}^{+ 2}$ &$ 1.3_{-0.5}^{+ 0.3}$ & $ 37_{ -4}^{+ 4}$ & $ 68_{ -5}^{+ 9}$ & $ 0.87_{- 0.11}^{+ 0.05}$ & $42.5_{- 0.1}^{+ 0.8}$ & $0.023_{-0.002}^{+0.002}$ & $ 0.5_{- 0.1}^{+ 0.1}$ & $ 0.4_{- 0.1}^{+ 0.1}$\\
NGC~3281 & $ 68_{ -2}^{+ 1}$ & $ 19_{ -2}^{+ 3}$ & $ 14_{ -1}^{+ 1}$ &$ 0.4_{-0.2}^{+ 0.2}$ & $ 38_{ -4}^{+ 3}$ & $ 19_{ -6}^{+ 6}$ & $ 0.99_{- 0.01}^{+ 0.01}$ & $44.2_{- 0.1}^{+ 0.9}$ & $0.151_{-0.008}^{+0.010}$ & $ 2.9_{- 0.4}^{+ 0.6}$ & $ 2.7_{- 0.4}^{+ 0.5}$\\
Cen~A & $ 50_{ -9}^{+ 10}$ & $ 17_{ -3}^{+ 3}$ & $ 10_{ -2}^{+ 2}$ &$ 0.3_{-0.2}^{+ 0.3}$ & $ 89_{ -13}^{+ 11}$ & $ 38_{ -9}^{+ 8}$ & $ 0.81_{- 0.16}^{+ 0.10}$ & $42.5_{- 0.1}^{+ 0.8}$ & $0.021_{-0.002}^{+0.002}$ & $ 0.4_{- 0.1}^{+ 0.1}$ & $ 0.3_{- 0.1}^{+ 0.1}$\\
NGC~5135 & $ 63_{ -5}^{+ 3}$ & $ 17_{ -2}^{+ 5}$ & $ 12_{ -2}^{+ 1}$ &$ 0.4_{-0.3}^{+ 0.4}$ & $ 71_{ -6}^{+ 5}$ & $ 17_{ -10}^{+ 10}$ & $ 0.97_{- 0.04}^{+ 0.01}$ & $43.6_{- 0.1}^{+ 0.8}$ & $0.079_{-0.006}^{+0.007}$ & $ 1.4_{- 0.3}^{+ 0.4}$ & $ 1.2_{- 0.3}^{+ 0.4}$\\
NGC~5643 & $ 62_{ -6}^{+ 4}$ & $ 14_{ -2}^{+ 4}$ & $ 13_{ -1}^{+ 1}$ &$ 0.8_{-0.5}^{+ 0.5}$ & $ 56_{ -9}^{+ 11}$ & $ 74_{ -12}^{+ 8}$ & $ 0.97_{- 0.04}^{+ 0.02}$ & $43.0_{- 0.1}^{+ 0.8}$ & $0.040_{-0.003}^{+0.004}$ & $ 0.6_{- 0.1}^{+ 0.2}$ & $ 0.5_{- 0.1}^{+ 0.2}$\\
NGC~5728 & $ 66_{ -3}^{+ 2}$ & $ 17_{ -1}^{+ 2}$ & $ 14_{ -1}^{+ 1}$ &$ 0.7_{-0.4}^{+ 0.4}$ & $ 48_{ -6}^{+ 7}$ & $ 80_{ -8}^{+ 5}$ & $ 0.99_{- 0.01}^{+ 0.01}$ & $43.4_{- 0.1}^{+ 0.9}$ & $0.063_{-0.004}^{+0.004}$ & $ 1.1_{- 0.1}^{+ 0.2}$ & $ 1.0_{- 0.1}^{+ 0.2}$\\
NGC~7172 & $ 69_{ -1}^{+ 1}$ & $ 29_{ -1}^{+ 1}$ & $ 14_{ -1}^{+ 1}$ &$ 0.0_{-0.0}^{+ 0.1}$ & $ 20_{ -1}^{+ 1}$ & $ 50_{ -3}^{+ 3}$ & $ 0.99_{- 0.01}^{+ 0.01}$ & $43.4_{- 0.1}^{+ 1.2}$ & $0.064_{-0.002}^{+0.002}$ & $ 1.9_{- 0.1}^{+ 0.1}$ & $ 1.8_{- 0.1}^{+ 0.1}$\\
\hline
\end{tabular}\\
Notes.--- Torus  model parameters derived from the fits with \B. 
Median values of each posterior distribution are listed with their corresponding $\pm1\sigma$ values around the median.
\end{center}
\end{table*}

\begin{table*}
\begin{center}
\scriptsize
\caption{Torus  model parameters from the global posterior distributions}
\begin{tabular}{lccccccccccccccc}
\hline
\hline
AGN &
\multicolumn{1}{c}{$\sigma_{\rm torus}$} &
\multicolumn{1}{c}{$Y$} &
\multicolumn{1}{c}{$N_0$} &
\multicolumn{1}{c}{$q$} &
\multicolumn{1}{c}{$\tau_{\rm V}$} &
\multicolumn{1}{c}{$i$} & 
\multicolumn{1}{c}{$C_{\rm T}$}  &
\multicolumn{1}{c}{$\log L_{\rm bol}^{(\rm mod)}$}   &
\multicolumn{1}{c}{$r_{\rm in}$} &
\multicolumn{1}{c}{$r_{\rm out}$} &
\multicolumn{1}{c}{$H$} 
\\
Type & [deg]  & & & & & [deg] & & [erg/s]  & [pc] & [pc] & [pc] \\
\hline
All & $  56_{ -22}^{+   6}$ & $  18_{  -6}^{+   2}$ & $  12_{  -4}^{+   1}$ & $ 0.5_{-0.5}^{+ 0.6}$ & $  81_{ -43}^{+  23}$ & $  45_{ -32}^{+  14}$ & $0.88_{-0.38}^{+0.06}$ & $43.3_{-0.4}^{+ 0.5}$ & $0.066_{-0.030}^{+0.048}$ & $ 1.2_{-0.8}^{+ 0.4}$ & $ 1.0_{-0.6}^{+ 0.3}$ \\ 
Type-1 & $  19_{  -3}^{+   3}$ & $  19_{ -10}^{+   3}$ & $  12_{  -2}^{+   0}$ & $ 0.7_{-0.5}^{+ 0.5}$ & $ 113_{ -31}^{+  20}$ & $  52_{ -48}^{+   9}$ & $0.18_{-0.06}^{+0.06}$ & $43.9_{-0.8}^{+ 0.3}$ & $0.144_{-0.090}^{+0.048}$ & $ 1.6_{-0.4}^{+ 0.4}$ & $ 1.0_{-0.6}^{+ 0.0}$ \\ 
HBLR & $  56_{  -8}^{+   4}$ & $  17_{  -8}^{+   2}$ & $  11_{  -4}^{+   1}$ & $ 0.8_{-0.8}^{+ 1.2}$ & $  98_{ -60}^{+  34}$ & $  43_{ -32}^{+   9}$ & $0.88_{-0.14}^{+0.04}$ & $43.7_{-0.7}^{+ 0.2}$ & $0.066_{-0.024}^{+0.054}$ & $ 1.2_{-0.4}^{+ 0.8}$ & $ 1.0_{-0.3}^{+ 0.6}$ \\ 
NHBLR & $  64_{ -11}^{+   2}$ & $  18_{  -3}^{+   2}$ & $  13_{  -3}^{+   0}$ & $ 0.4_{-0.4}^{+ 0.2}$ & $  43_{ -11}^{+  17}$ & $  48_{ -28}^{+  17}$ & $0.96_{-0.10}^{+0.02}$ & $43.2_{-0.8}^{+ 0.1}$ & $0.060_{-0.042}^{+0.006}$ & $ 0.8_{-0.4}^{+ 0.8}$ & $ 1.0_{-0.6}^{+ 0.3}$ \\
\hline
\end{tabular}\\
Notes.--- Torus parameters from the global posterior distributions of each subgroup. 
\end{center}
\end{table*}

\section{RESULTS and DISCUSSIONS}
\subsection{Infrared SEDs with \B\ Fitting}
The results of the fitting process to the IR SEDs are the posterior distributions for the six parameters that
describe the model (defined in Table~3), the foreground extinction and the multiplicative factor needed
to match the SED fluxes.
However, we can also translate the results into corresponding spectra, as shown in Figure~1.

Figure~1 shows the observed SEDs and nuclear MIR spectra (black filled dots) of the galaxies
with the best fit results overlaid (blue solid lines), based on the inference done with \B.
 The fitted models correspond to those described by the median of the posterior distribution of each parameter.
All the derived torus parameters obtained from \B\ are presented in Table~4.

Some SEDs show lower $Q$-band fluxes than those predicted by the fitted model.
This effect is prominent when silicate 9.7~$\mu$m feature is observed in deep absorption. 
Although this may suggest that the model spectra still have a difficulty
 to reproduce the 18~$\mu$m silicate feature, the difference is only within a factor of 3
  in the worst case (NGC~3281). See \cite{ram09}  for further discussion on the $Q$-band excess.
 
The SEDs of NGC~5506 and NGC~7172 show NIR excesses compared to the model spectra as shown in Figure~1.
NIR interferometric observations \citep{kis09, kis11} and NIR reverberation
 mapping \citep{kis07, kos09} of Seyfert galaxies suggest that the AGN torus has
  much smaller sublimation radius than expected from Eq.~(1). 
   \cite{kaw11} also showed in their model that if they add the rim darkening effect of the accretion disk,
the torus inner radius naturally connects to the outer disk. 
Thus, the NIR excess is readily accounted for by this connection.
This is also shown in \cite{sta12} when they apply the rim darkening effect to produce the model SEDs.
However,  in this study, we did not include the SED of a hot dust component or apply such torus geometry including
 rim darkening effect for the fitting as this hot dust remains rather unconstrained. 
 Further discussion on the possible origins of NIR excesses can be found in \cite{alo11}.

\subsection{General Torus Properties for Whole Sample}

In this section, we describe how we obtained the global distribution of the
 torus model parameters for the whole sample and the three subgroups considered here.
 To take full advantage of the data
 employed here, we apply a hierarchical Bayesian approach to derive information about the
 global distribution of the \textit{CLUMPY} parameters for a given subgroup. We use a generalized beta distribution
 as the prior for each parameter (given that they are defined in closed intervals) and learn
 the hyperparameters of the prior using importance sampling (e.g. \citealt{bre14}).
 This allows us to derive the posterior distribution for each parameter taking into account all
 the observed data that belongs to an AGN subgroup. 
 
 In Table 5 we report the median parameters of the global posterior distributions for the 
 whole sample, as well as for each subgroup. Note that the global posterior distribution of each
 parameter can be derived only in the circumstance that the prior distribution is the same
 for all the sources. This is not the case for the inclination angle because we use the 
 constrained prior distributions for some sources as described in Section 3. For the inclination 
 angle, we derive the median parameters of the individual galaxy fits within the each subgroup.

Based on interferometric observations, \cite{kis11} reported a typical torus half-light radius of $\sim1$~pc for local AGN at MIR wavelengths.
Our derived torus outer radius $r_{\rm out} = 1.2_{-0.8}^{+0.4}$~pc is consistent with the interferometry results, although a little larger.
This is understood because  here we are considering colder dust within the torus than that traced by the MIR interferometry, as we include 
data at $\lambda > 20$~$\mu$m in our fits.
These colder clumps will be generally located at larger radii, which explain the value of $r_{\rm out}$ obtained from the global posterior distribution.
The torus outer radius including cooler dust would be larger still than the value of $r_{\rm out} \sim 1.2$~pc.
Further studies including \textit{Herschel}, SOFIA, and/or ALMA observations will help constraining the extent of the FIR-emitting dust.
Indeed, some pilot studies already showed the importance of FIR data to trace cooler dust \citep{ram11b,gar14}.

\begin{table*}
\begin{center}
\scriptsize
\caption{Results of KLD test for each parameter among each subgroup}
\begin{tabular}{lccccccccccccccc}
\hline
\hline
AGN Type &
 \multicolumn{1}{c}{$\sigma_{\rm torus}$} &
\multicolumn{1}{c}{$Y$} &
\multicolumn{1}{c}{$N_0$} &
\multicolumn{1}{c}{$q$} &
\multicolumn{1}{c}{$\tau_{\rm V}$} &
\\
\hline
Type-1 vs HBLR   &{\bf  3.13 }&{\bf  1.26 }& 0.10 & 0.42 & 0.10\\
Type-1 vs NHBLR  &{\bf  2.22 }& 0.22 & 0.05 & 0.20 &{\bf  4.83}\\
HBLR   vs NHBLR  &{\bf  1.78 }&{\bf  1.63 }& 0.24 & 0.53 &{\bf  1.56}\\
\hline
\end{tabular}\\
Notes.--- KLD is calculated for the global posterior distribution of each parameter among the subgroups. Values larger than 1 are shown in bold.
\end{center}
\end{table*}

\begin{figure*}
\begin{center}
\includegraphics[width=6.0cm]{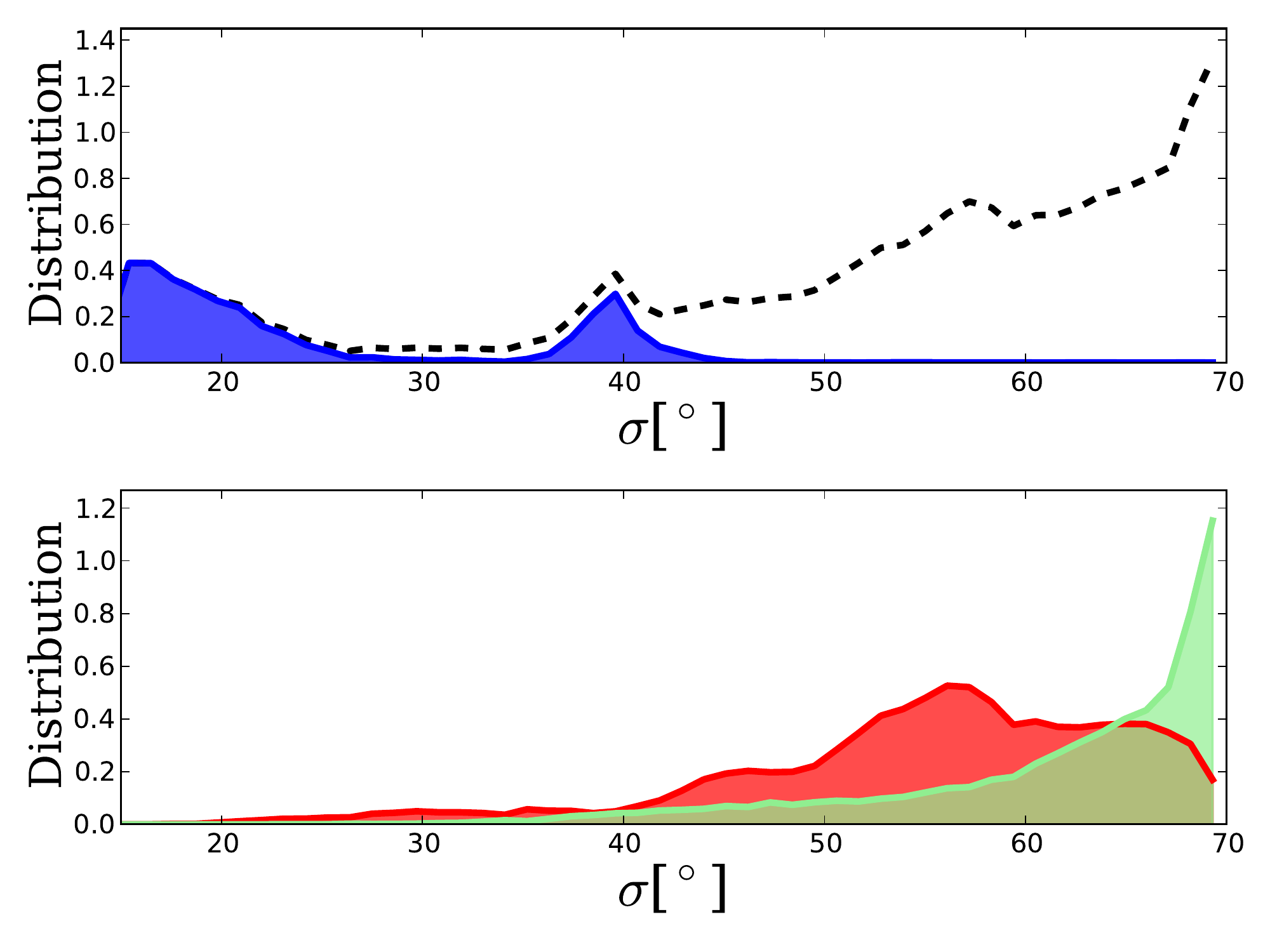}~
\includegraphics[width=6.0cm]{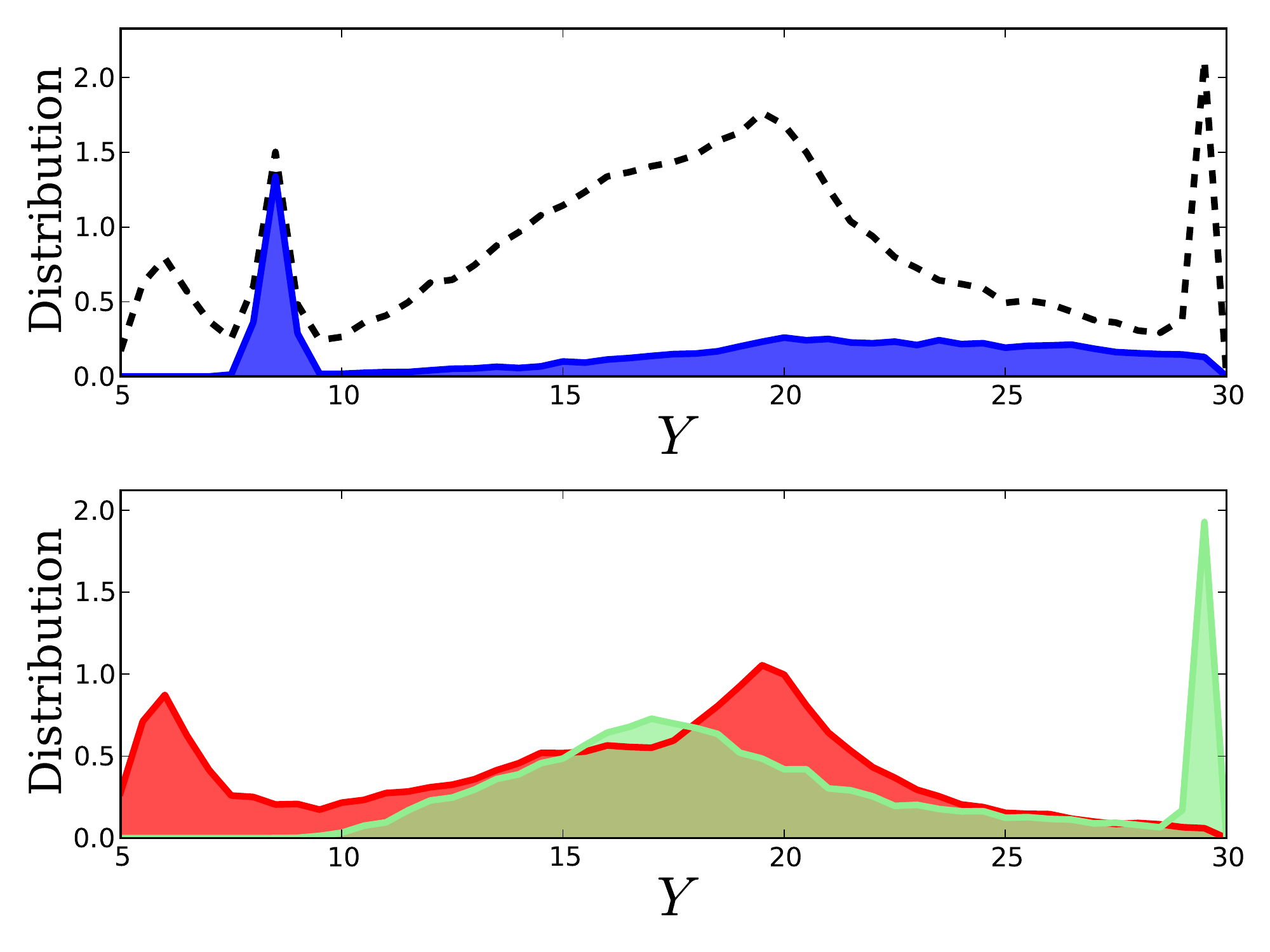}~
\includegraphics[width=6.0cm]{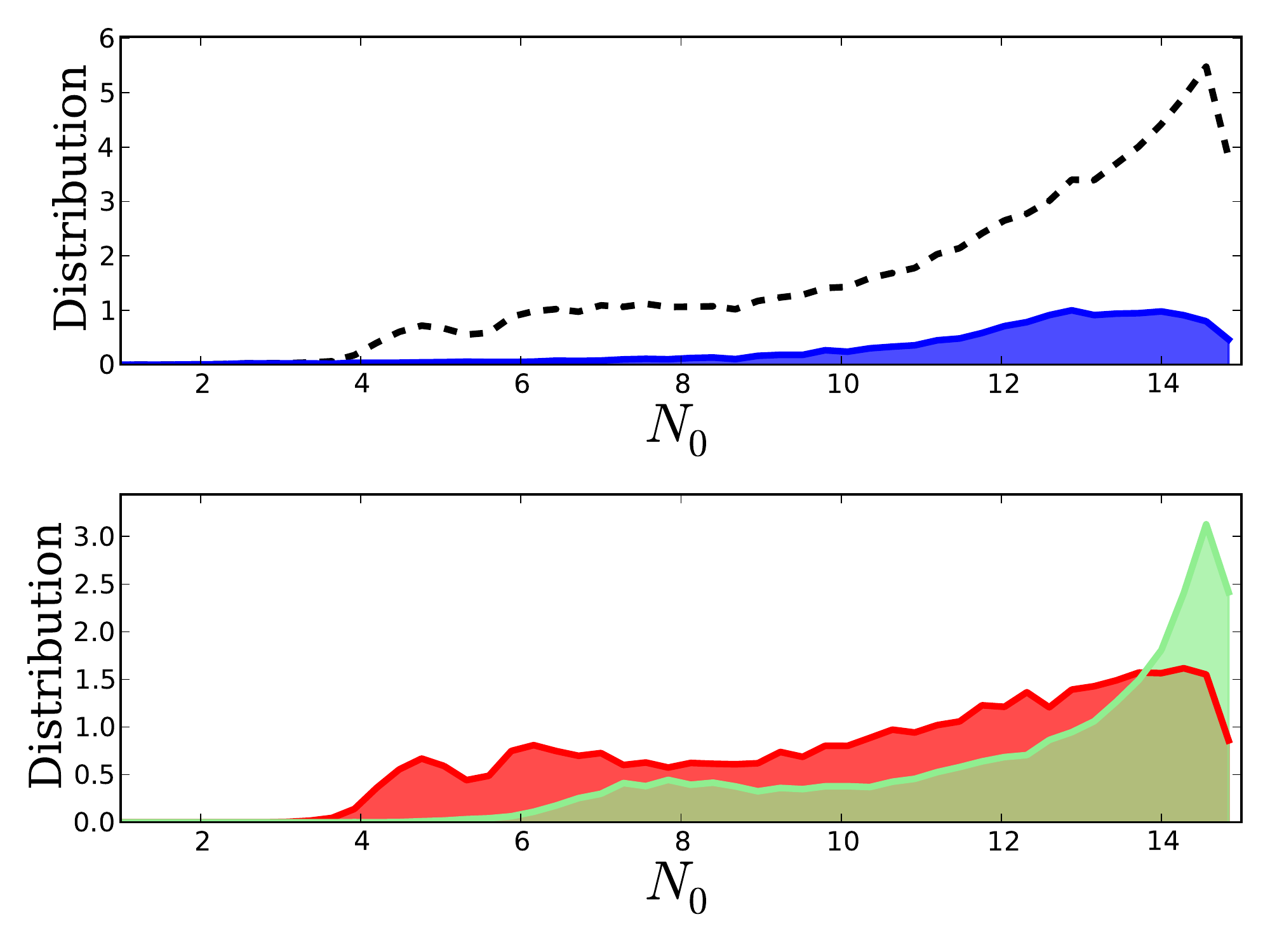}\\
\includegraphics[width=6.0cm]{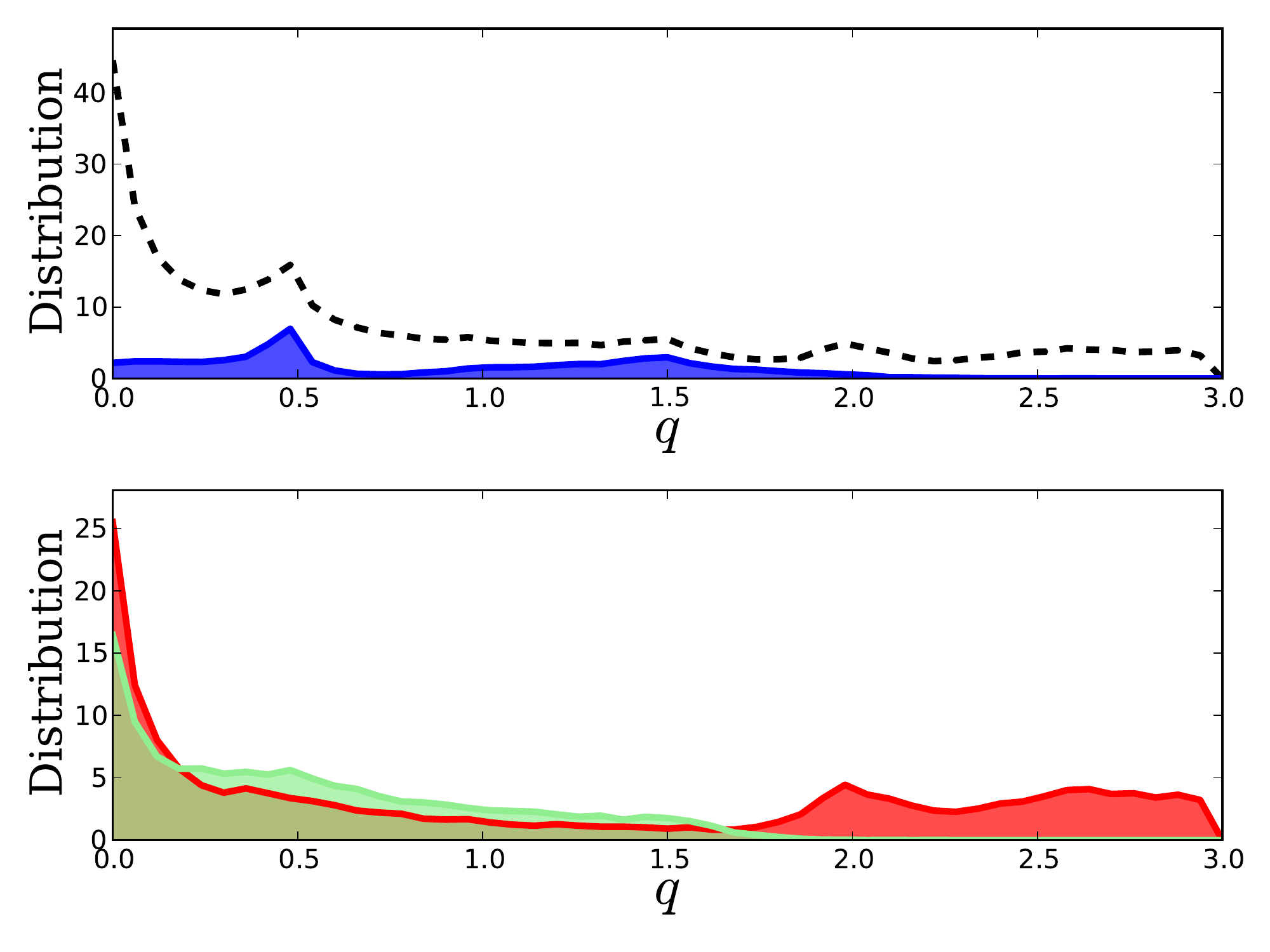}~
\includegraphics[width=6.0cm]{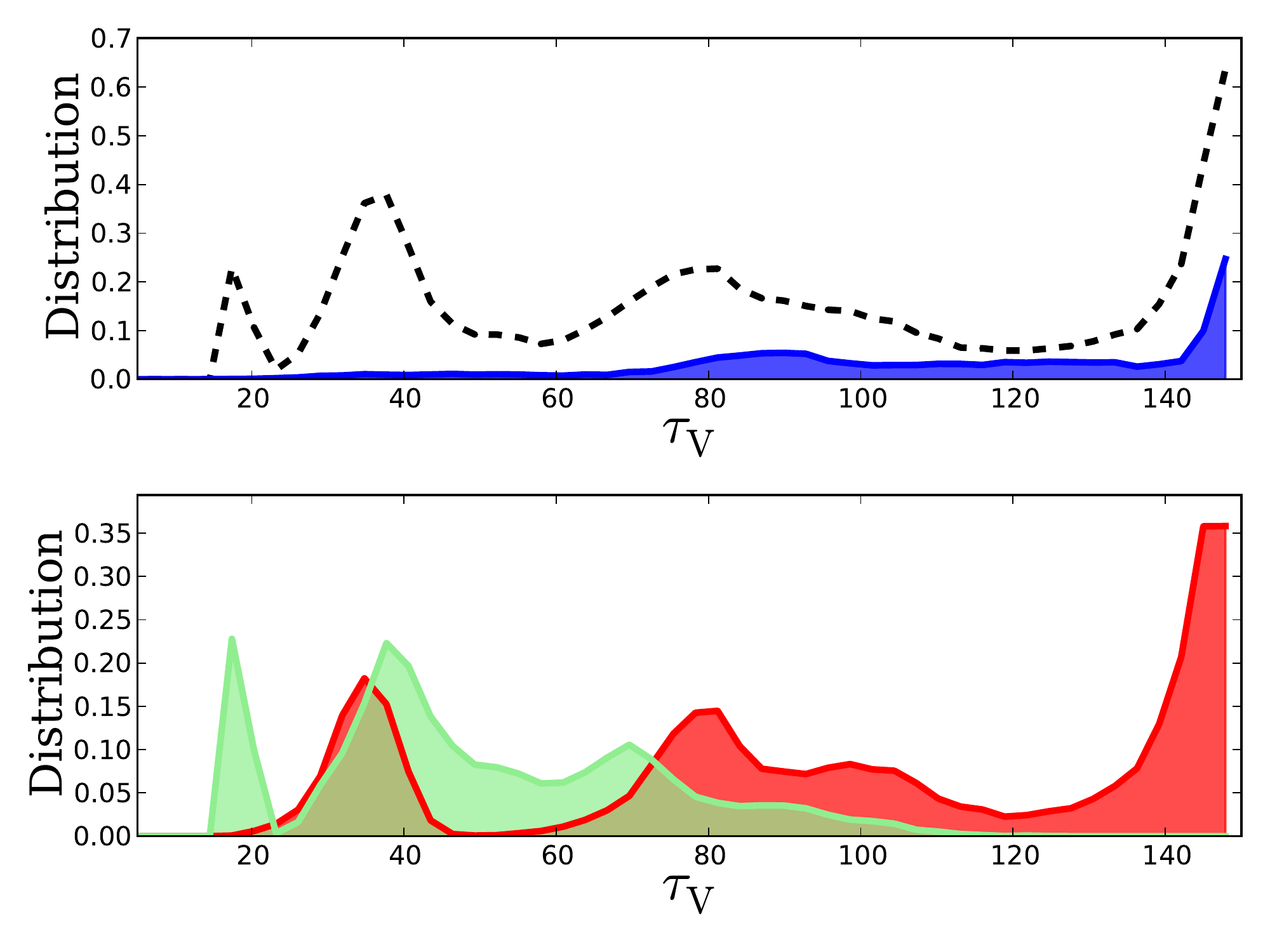}
\caption{Histograms of each physical parameter discussed in Section~4.3. 
Top panel of each figure represents the histogram of the whole sample. 
Blue/red/green filled color represents the histogram of Type-1/HBLR/NHBLR, respectively.}
\end{center}
\end{figure*}

\begin{figure*}
\begin{center}
\includegraphics[width=6.0cm]{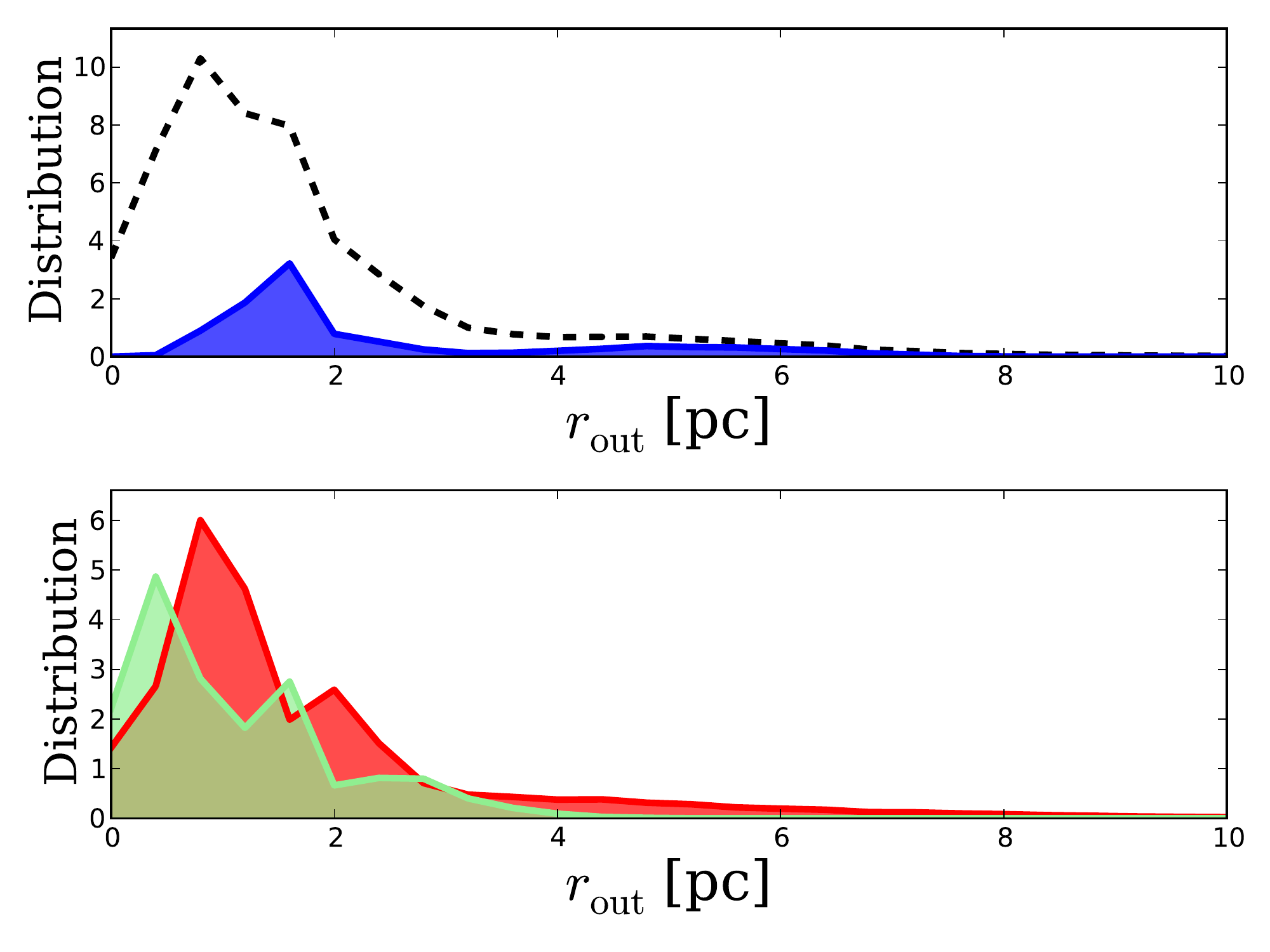}~
\includegraphics[width=6.0cm]{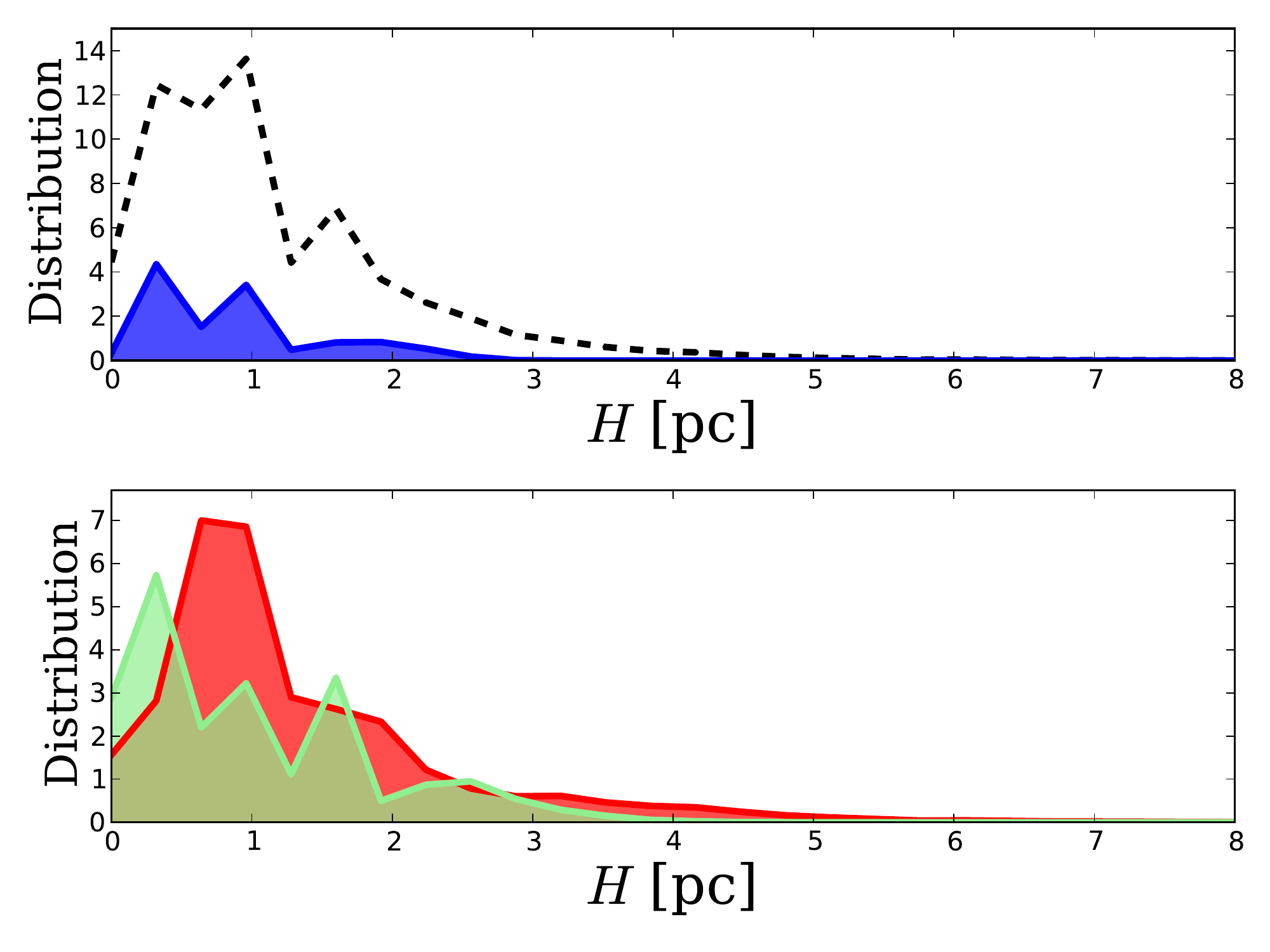}~
\includegraphics[width=6.0cm]{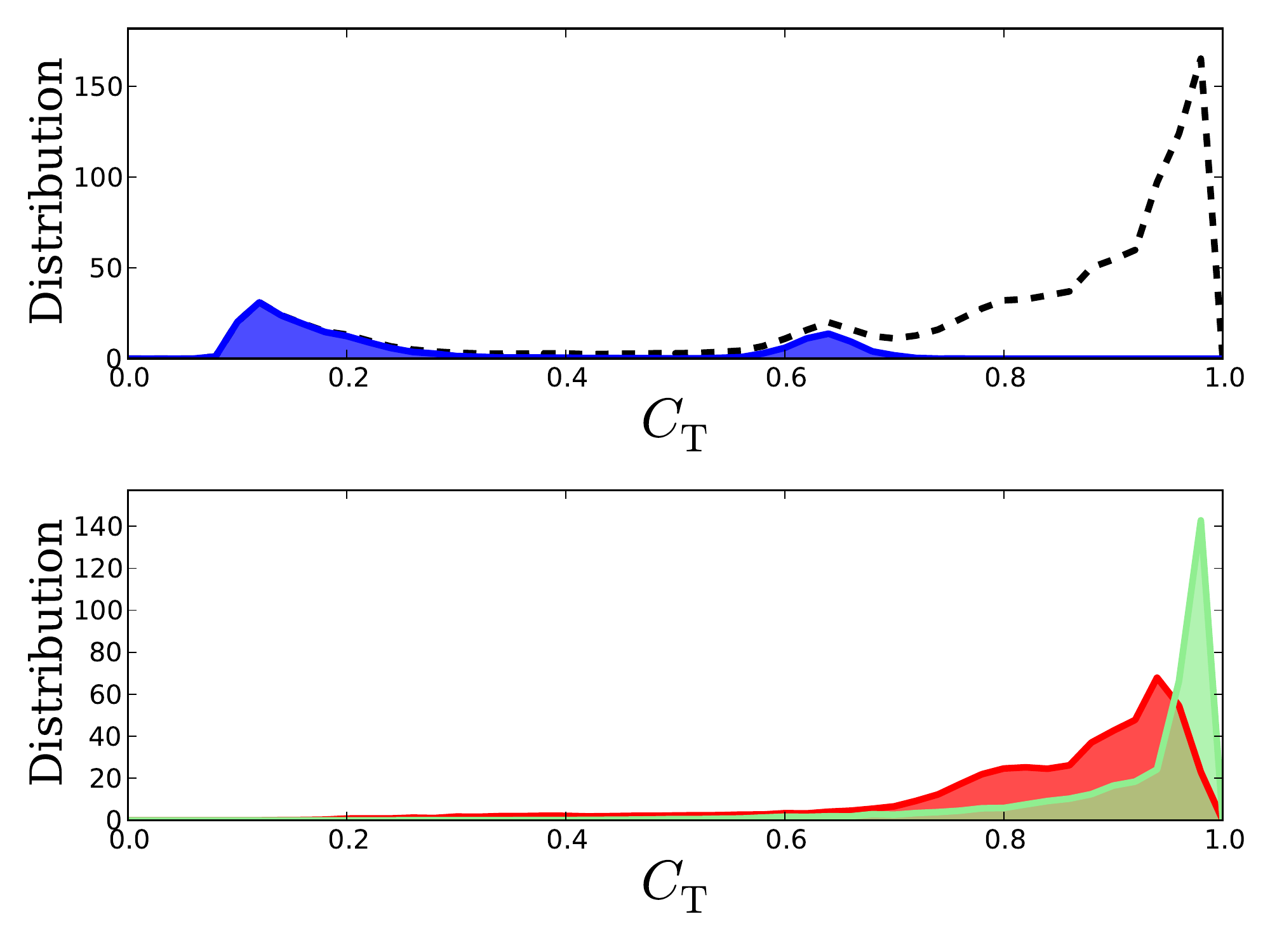}
\caption{Same as in Figure~1, but for the $r_{\rm out}$, $H$, and $C_{\rm T}$ parameters.}
\end{center}
\end{figure*}

\subsection{Distribution of Torus Model Parameters}
The key science direction of this paper is to investigate the torus morphology quantitatively
 by deriving torus parameters for each subgroup and comparing them. 
 The median values of the model parameters fitted to our nuclear IR SEDs are reported in Table~4.
  The median values of the global posterior distributions of the free parameters 
  $\sigma$, $Y$, $N_0$, $q$, and $\tau_{\rm V}$ for each subgroup are compiled in Table~5.
  As discussed in Section 4.2, for the inclination angle $i$ we derive the median values from 
  the individual galaxy fits for each subgroup.

Figure~2 shows the global posterior distributions of each physical parameter for the different subgroups.
Black, blue, red, and green histograms show the parameter distribution of
 all, type-1, HBLR, and NHBLR sources, respectively.

As shown in Figure~2, for some of the parameters, 
the global distributions are similar for the three subgroups, 
but others are clearly different.
In order to quantify these differences, we follow the same approach as in \cite{ram11}. 
They used the Kullback-Leibler divergence \citep[KLD;][]{kul51}
to show that the joint posterior distributions of type-1 and type-2 AGN were quantitatively different.
The KLD takes into account the full shape of two posterior distributions to compare them.
When the two distributions are identical, the value is ${\rm KLD}=0$ and, 
the larger the KLD value, the more different the two distributions.
\cite{ram11} concluded that if ${\rm KLD}>1.0$,
the two posteriors can be considered significantly different. 
We calculate the KLD values for the global distributions
 of each torus parameter among the three groups. 
 The values are reported in Table~6.

We find significant differences for the parameters $\sigma$, $Y$, and $N_{\rm 0}$ between type-1 AGN and HBLR AGN.
The differences in these parameters between type-1 and type-2 AGN were already reported in \cite{ram11} with larger significance,
but based on fits to NIR and MIR photometry only, where spectroscopic data were not included.
Besides, they did not consider information from spectropolarimetry data, as we are doing here.
Therefore, we confirm the results of \cite{ram11} after including $N$~band spectroscopy to the IR photometry, 
which is crucial to constrain the six torus parameters \citep{alo11,ram14}.
We also find that the parameters $\sigma$, $Y$, and $N_{\rm 0}$ of HBLR and NHBLR AGN are significantly different.
Considering the average values in Table~5, the tori of NHBLR AGN have larger $\sigma$, larger $Y$, and larger $N_{\rm 0}$ than those of HBLR AGN.

There are various possible interpretations for the difference in $\sigma$ among the subgroups. 
One possibility is that the smaller $\sigma$ could be due to larger AGN luminosities \citep[receding torus model;][]{law91, ric13}. 
However, as shown in Table~5, the differences of median bolometric AGN luminosities among the subgroups are within the uncertainties.
Therefore, we consider that the effect of the AGN luminosity is negligible in this study. 
Another possible interpretation is a selection bias in the optical type-1/type-2 AGN selection. 
\cite{ram11} and \cite{eli12} discussed that AGN classification would depend on the distribution of the obscuring material; 
type-1 AGN would be preferentially selected from lower-obscuration AGN, while type-2 AGN (HBLR and NHBLR) 
from higher-obscuration AGN. 
This could be partly producing the differences in $\sigma$ that we found for type-1 and type-2 AGN, but correcting this effect 
quantitatively is extremely difficult and beyond the scope of this paper.

  \subsection{Distribution of Covering Factor}
  
As shown in Section 3.1., we can derive physical parameters of the torus model by combining the model parameters. 
  The individual values of these physical parameters are reported in Table~4, and those obtained from the global 
  posterior distribution for each subgroup are shown in Figure~3 and Table~5.
  
An interesting comparison can be made between the geometrical covering factor of the torus model
 ($C_{\rm T}$; described in equation 6) of the different subgroups and the average column densities derived from X-ray data ($N_{\rm H}$).
NHBLRs have the largest column densities, with an average value of $\log N_{\rm H} \sim 24.0$~cm$^{-2}$ (i.e. 
Compton-thick), followed by HBLRs, with $\log N_{\rm H} \sim 23.4$~cm$^{-2}$, and type-1s, with $\log N_{\rm H} \sim 21.8$~cm$^{-2}$.  
Based on hard X-ray (50--200 keV) observations of nearby AGN obtained with \textit{INTEGRAL}, 
\cite{ric11} reported differences between the X-ray reflection component of type-1 and type-2 AGN. 
Type-1 and ``lightly obscured'' AGN with $N_{\rm H} \le 10^{23}$~cm$^{-2}$ have the
same X-ray reflection component, with reflection amplitude $R \sim 0.4$. 
On the other hand, ``mildly obscured'' AGN ($N_{\rm H} \ge 10^{23}$~cm$^{-2}$) show
a clearly stronger X-ray reflection component with $R \sim 2.2$, suggesting that the central engine of ``mildly obscured'' AGN
would be covered by an X-ray reflection wall.
Our results are in good agreement with \cite{ric11} if we consider the $C_{\rm T}$ and $N_{\rm H}$ values for each subgroup.
The type-1 AGN in our sample fall in the ``lightly obscured'' AGN in their study, and indeed they 
show small covering factors ($C_{\rm T} \sim 0.18$), suggesting small torus X-ray reflection solid angle. 
The HBLR and NHBLR AGN subgroups would fall in the ``mildly obscured'' AGN category, and we found 
large covering factors for them ($C_{\rm T} \sim 0.88$ and 0.96 respectively), suggesting a larger X-ray reflection component 
\citep[see also][]{ric14}. Figure~4 shows a schematic illustration of the torus geometrical differences among type-1 (top), 
HBLR (middle), and NHBLR (bottom).

\begin{figure}
\begin{center}
\includegraphics[width=8.2cm]{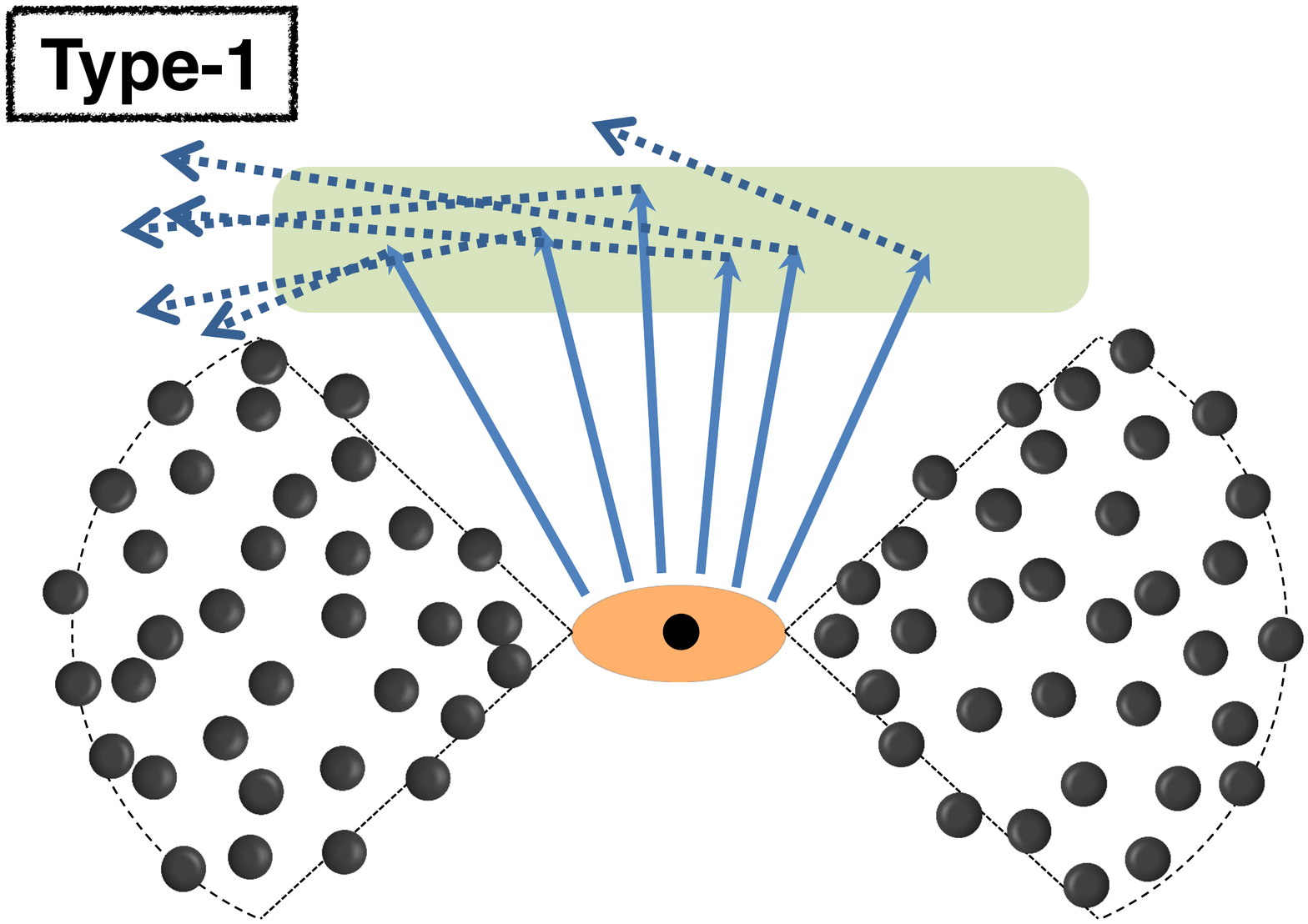}\\
\includegraphics[width=7.3cm]{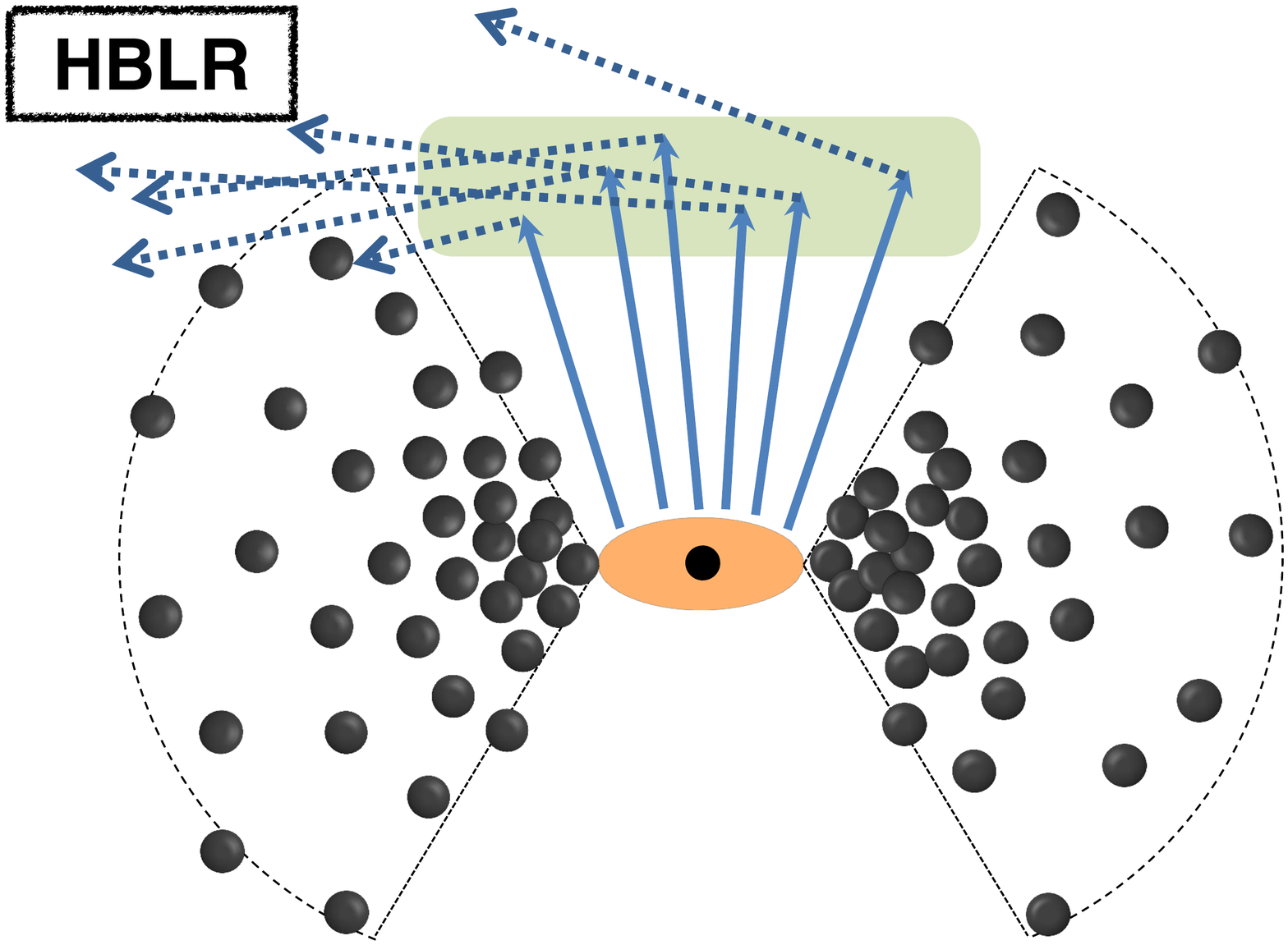}\\
\includegraphics[width=7.5cm]{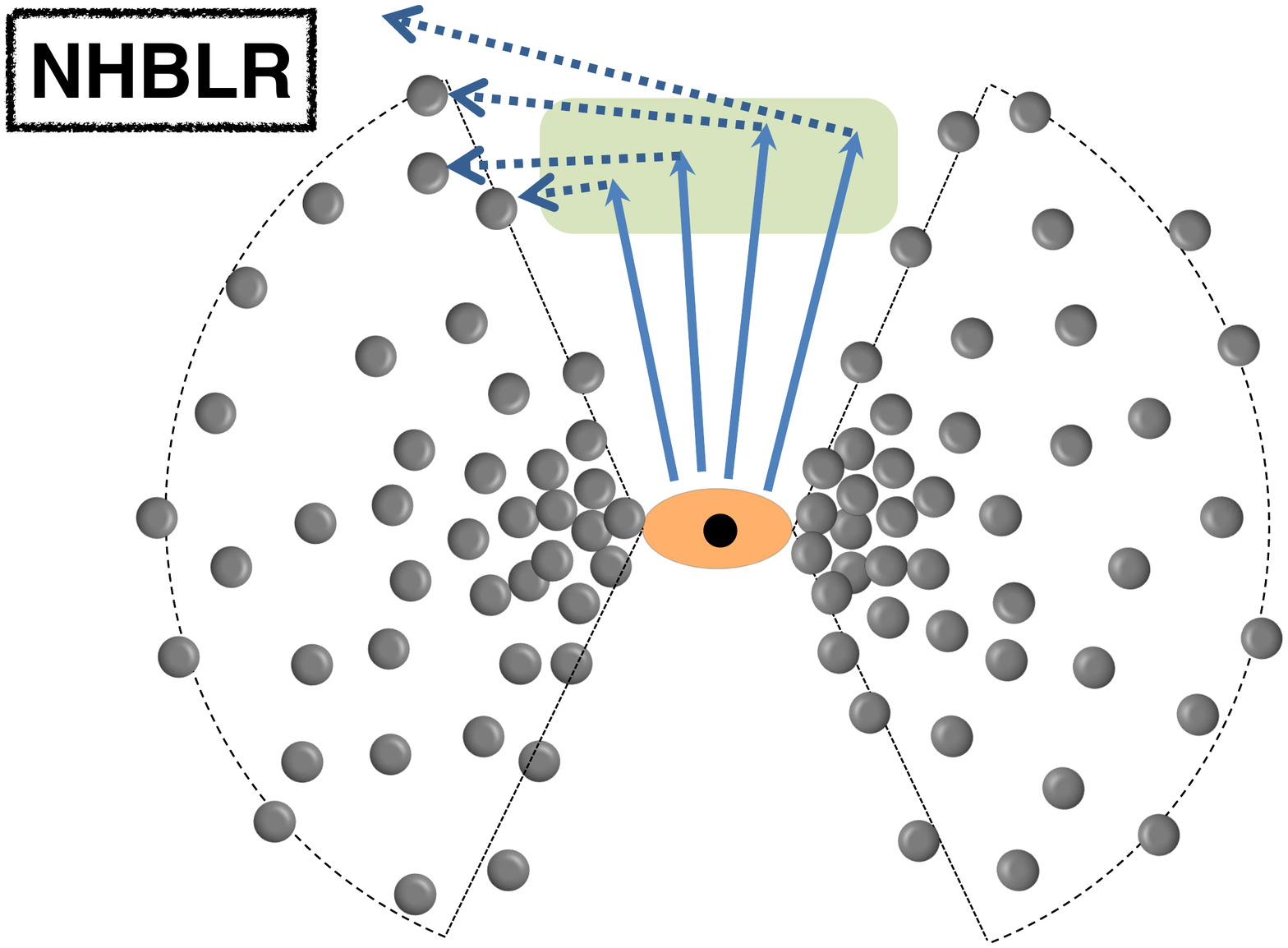}
\caption{
 Schematic illustration of the torus geometry for type-1 AGN (top), 
 HBLR (middle), and NHBLR (bottom). 
 The difference in color intensity between the two bottom panels 
 shows the difference in optical depth of the clumps  $\tau_{\rm V}$,
  where darker color is larger  $\tau_{\rm V}$. 
  The orange region represents the BLR. 
  The green area represents the media where some of the incoming 
  BLR emission is scattered and then polarized
  (in all diagrams we describe only scattering from the polar 
  scattering region, and ignore the inner equatorial scattering region, 
  dominantly responsible for the polarized flux in Type 1 objects).
  The blue solid arrows represent the path of BLR photons and 
  the blue dashed arrows represent the path of the polarized BLR photons.
  The observer is assumed to be on the left side of the torus
  with an inclination angle of 52$^{\circ}$, 43$^{\circ}$, and 48$^{\circ}$, respectively (see Table~5).
  The only photons scattered along these lines of sight are shown.
}
\end{center}
\end{figure}

\begin{figure}
\begin{center}
\includegraphics[width=8.5cm]{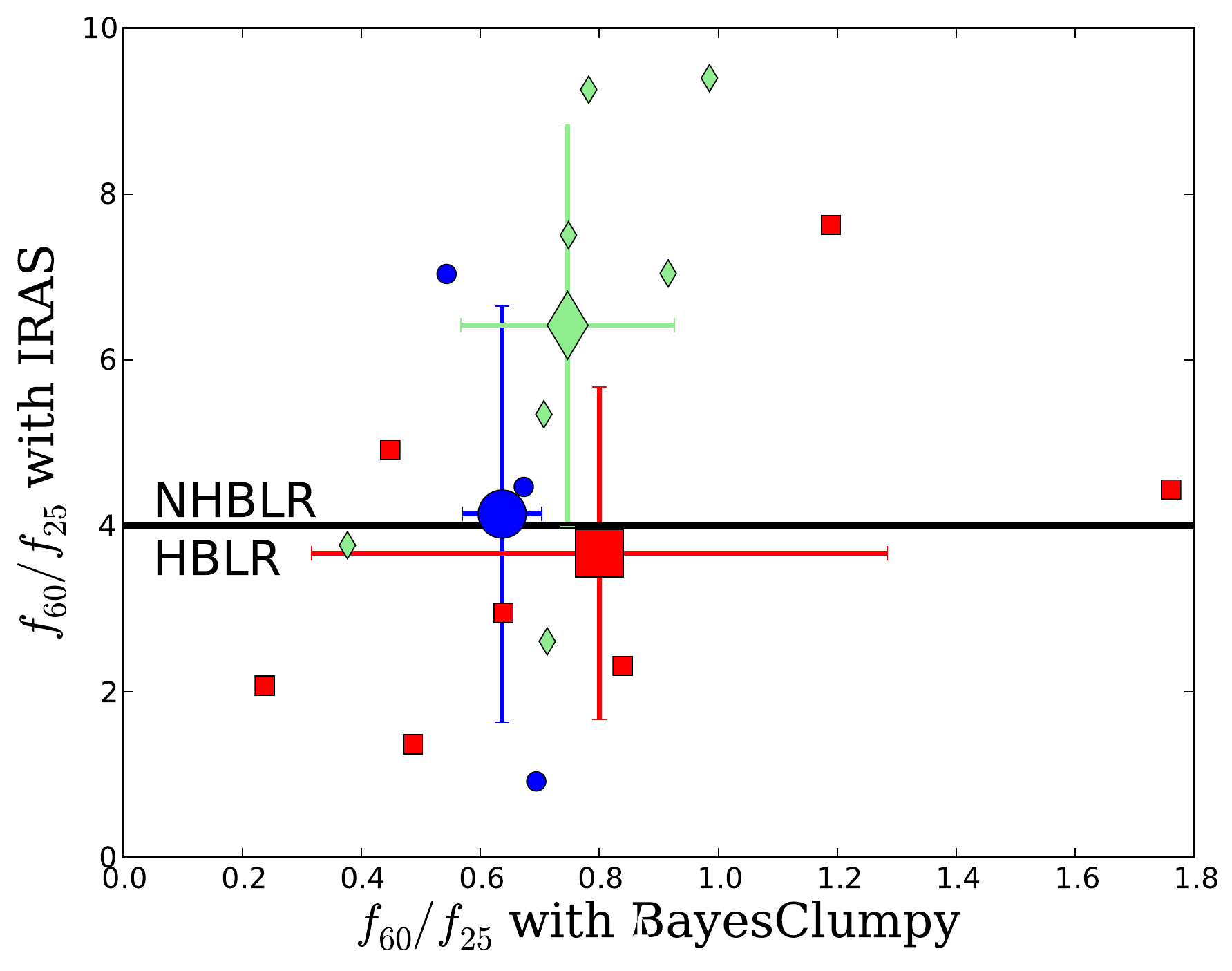}
\caption{
Plot comparing the $f_{60}/f_{25}$ colors measured from \textit{IRAS} photometry and extrapolated from the clumpy torus model fits.
Blue circle/red square/green diamond represents type-1, HBLR, and NHBLR AGN, respectively. The larger symbol
represents the average value of each subgroup. The horizontal line at $f_{60}/f_{25}=4$ 
is the boundary between HBLR and NHBLR AGN 
according to \cite{hei97}, where HBLR AGN have $f_{60}/f_{25}<4$ and NHBLR AGN $f_{60}/f_{25}>4$.
} 
\end{center}
\end{figure}

\subsection{Torus Model Morphological Differences between HBLR and NHBLR}

In this section we focus on the differences between the modeled tori of HBLR and NHBLR. 
 In the case of HBLR, we obtain smaller $\sigma$ values than for NHBLR,
  which is equivalent to larger torus opening angles ($90^{\circ}-\sigma$). 
  This implies that HBLR objects can have a larger scattering region.
 (see middle panel of Figure~4).
  The scattering region (shown schematically as a filled green bar) 
  can be larger due to the larger opening angle 
  of the torus, allowing more photons to be scattered, and hence polarized, 
  from the BLR.  
  We note that the larger opening of the ionization cone, if unresolved, will produce 
  a slightly lower degree of polarization due to (partial) cancellation of those
   polarization vectors at the edges of the scattering region.
   However, in our case, the increased amount of scattered photons will 
   significantly increase the polarized flux, with only a very small reduction
    in the degree of polarization.  
To confirm this effect, we produced a toy polarization model assuming that 
the scattering region is a two dimensional biconical structure centered to the central engine.
Then, we measure the degree of polarization and polarized flux, and
 found that the measured polarized flux is larger for HBLR than those for NHBLR, supporting
 our results.
     \cite{mil90} also find the same results using a 3 dimensional cone and 
     more sophisticated modeling, and more complex assumptions.

  In the case of the NHBLR, we obtain larger 
  $\sigma$ values than for HBLR.
   This means that the probability that scattered radiation from the BLR can be blocked is higher than for HBLR objects.
This is also in agreement with the larger value of $\log N_{\rm H} \sim 24.0$~cm$^{-2}$ estimated from X-ray observations of 
the NHBLR objects in our sample.

     To summarize, as shown in the bottom panel of Figure~4, the chance to observe scattered 
     (polarized) flux from the BLR is reduced by the double effect of
      (a) less scattering of the flux from the BLR (due to the reduced scattering area) and
       (b) more obscuration between the observer and the scattering region.  
Therefore, the classification of an AGN as HBLR or NHBLR is probabilistic, and it would depend
on the intrinsic properties of the torus, in particular of $\sigma$. 
       This could be a reasonable explanation for the lack of a hidden (polarized) BLR in some type-2 objects\footnote{
       A similar explanation for the lack of HBLR detection in $\sim$40\% of type-2 objects, 
       based on the distribution of dust within the torus and its inclination being not as simple
        as predicted by the unified model, was shown in the talk by C. Ramos Almeida at the Polarization
        \& active galactic nuclei workshop held 2012 October 16-17 at the Royal Observatory of Belgium.}. 

       	However, we note that the classification of the galaxies as HBLR and NHBLR is mainly based on 
	spectropolarimetric observations from 3--4 m telescopes, and some of the NHBLR could be then misclassified.
        Therefore, further higher sensitivity spectropolarimetry observations of 
       NHBLR AGN with 8~m class telescopes such as Subaru/FOCAS and/or VLT/FORS2 are highly encouraged
       to search for the HBLR in those AGN (Ramos Almeida et al. in prep.).
       
\subsection{Inclination Angle Effect on Detectability of HBLR in Type-2 AGN}

The dependence of HBLR detection in type-2 AGN on torus inclination angle is still a matter of debate.
This idea arises from the observational trend that the \textit{IRAS} 60~$\mu$m to 25~$\mu$m color ratio ($f_{60}/f_{25}$) of HBLR AGN
is $f_{60}/f_{25}<4$ on average while that of NHBLR AGN is $f_{60}/f_{25}>4$ \citep{hei97,lum01}.
Several authors have suggested that this trend is due to the inclination angle of the torus: 
the cooler and outer dust within the torus blocks the warm and inner hot dust
for edge-on views, producing the high $f_{60}/f_{25}$, while the warm inner dust can be seen from the face-on views, 
reducing the value of $f_{60}/f_{25}$.
Following this idea, type-2 AGN with low $f_{60}/f_{25}$ would tend to be detected as HBLR AGN 
due to the more face-on view of the torus and vice versa.

Here we can take advantage of the torus models fitted to our SEDs, which are available for each AGN and shown in 
Figure~1. The wavelength range covered by the models allows us to calculate $f_{60}/f_{25}$ color ratios for each source. 
We have also compiled $f_{60}/f_{25}$ color ratios from \textit{IRAS} for comparison, which probe much larger scales than 
our torus SEDs.
Out of 21 sources, we obtained 17 \textit{IRAS} $f_{60}/f_{25}$ colors with good quality of fluxes (${\rm FQUAL}=3$, 
which is the highest quality)\footnote{
See \cite{bei88} for the definition of FQUAL in the \textit{IRAS} catalogs. False detections may be included when ${\rm FQUAL}<3$.}.

Figure~5 shows the relationship between the $f_{60}/f_{25}$ flux ratios 
obtained from \textit{IRAS} and those from the torus model SEDs.
The averages \textit{IRAS} $f_{60}/f_{25}$ flux ratio for type-1, HBLR, and NHBLR AGN are
 $4.1 \pm 2.5$, $3.7 \pm 2.0$ and $6.4 \pm 2.4$, respectively,
  showing the previously mentioned correlation for HBLR and NHBLR AGN, although with large error bars.
However,  when we compare the values obtained from the torus model SEDs, which exclude contamination 
from the host galaxy, we find that they are very similar for the three groups and smaller than the \textit{IRAS} colors
($f_{60}/f_{25}$ = $0.63 \pm 0.07$, $0.80 \pm 0.48$, and $0.75 \pm 0.18$ for type-1, HBLR, and NHBLR AGN respectively).

These results show that 
the differences of \textit{IRAS} $f_{60}/f_{25}$ among the three subgroups are not produced from the torus dust. 
A similar result was reported by \cite{ale01}, but using X-ray observations.
Although the standard deviations of the average values of the \textit{IRAS} colors are large for the three subgroups, 
one possible explanation for the cooler \textit{IRAS} colors of NHBLR in comparison with those of HBLR AGN could be  
dust emission from stronger starbursts in their host galaxies.
This is in good agreement with previous results showing that highly obscured AGN tend 
to have higher star formation activity in their host galaxies
\citep{gou12,ich12,ich12b,ich14a,cas14}. Therefore, larger obscuration from the torus in NHBLR AGN (as shown 
in Figure~4) and higher star formation activity in the host galaxy could be somehow coupled.
For example, using three-dimensional hydrodynamical simulations, \cite{wad02} showed that starbursts and 
supernovae within the central 100~pc of host galaxies help lifting up the torus, 
suggesting that high star formation activity could influence the scale height of the torus. 
Comparing the nuclear and overall star formation activity of AGN with the torus obscuration 
is crucial to find out if they are coupled \citep[e.g.,][]{ima11, esq14,ich14a}.

\section{CONCLUSIONS}
We constructed 21 infrared torus-dominated SEDs including high spatial resolution 
NIR and MIR photometry, MIR spectroscopy, and \textit{Spitzer} and \textit{Herschel} FIR fluxes.
By performing SED fitting using clumpy torus models and a Bayesian approach we derived torus parameters
such as the torus covering factor ($C_{\rm T}$), the torus inner and outer radius ($r_{\rm in}$ and $r_{\rm out}$), 
and the torus scale height ($H$).
We divided the sample into subgroups based on whether or not  they are optically type-1, 
type-2 with observational hidden broad line regions sign (HBLR), 
and type-2 without any observational broad line region signs (NHBLR). 
Our results are summarized as follows:

\begin{enumerate}
\item Under the assumption of a clumpy distribution of the dust, we obtained a
 quantitative description of the torus geometry and intrinsic properties. 
 We found that the median torus outer radius for the whole sample $r_{\rm out}=1.2$~pc is consistent with
the results from MIR interferometry observations.

\item We found that the tori of Type-1 AGN have smaller $\sigma$, $Y$, 
$N_{\rm H}$, and $C_{\rm T}$ than those of HBLR and NHBLR. 
Moreover, the tori of NHBLR are thicker 
and therefore have higher $C_{\rm T}$ than those of HBLR.
These differences in the torus properties of HBLR and NHBLR
 AGN would make it more difficult to detect hidden BLR in NHBLR. 
 
 \item Combining $f_{60}/f_{25}$  colors obtained from \textit{IRAS} photometry and from torus model SEDs, we showed 
 that the low $f_{60}/f_{25}$ measured fof HBLR using \textit{IRAS} data are not due to a more face-on inclination of torus, but rather to star formation activity
in their host galaxies.

\end{enumerate}



\acknowledgments
We are grateful for useful comments from the anonymous referee.
We also thank C. Ricci, M. Stalevski for useful comments and discussions.
K.I. thanks the Department of Physics and Astronomy at University of Texas at San Antonio, where most of the research was conducted.
This paper is based on observations obtained at the Gemini Observatory, which is operated by the Association of Universities for Research in Astronomy, Inc., under a cooperative agreement with the NSF on behalf of the Gemini partnership.
This work was partly supported by the Grant-in-Aid for JSPS Fellows for young researchers (K.I.). 
C.P. acknowledges support from UTSA and NSF (grant number 0904421).
C.R.A. is supported by a Marie Curie Intra European Fellowship within the 7th European Community Framework Programme (PIEF-GA-2012-327934).
A.A.-H. acknowledges support from the Spanish Plan Nacional de Astronomia y Astrofisica under grant AYA2012-31447.

\end{document}